%% file: paper.tex
\pdfoutput=1

\def\isReadyToSubmit{1}   

\documentclass[letterpaper,twocolumn,10pt]{article}
\usepackage{etex}
\usepackage{usenix,epsfig,endnotes}
\usepackage[noadjust]{cite}
\usepackage[utf8]{inputenc}
\usepackage{amsopn, amssymb, amsmath, amsthm}
\usepackage[usenames,dvipsnames]{xcolor}
\usepackage{setspace}
\usepackage{algorithm}
\usepackage{algorithmicx}
\usepackage{algpseudocode}
\usepackage{tikz,pgfplots,pgfplotstable}
\usepgfplotslibrary{groupplots}
\usepackage{bm}
\usepackage{booktabs}
\usepackage{url}
\usepackage{array}
\usepackage{bbm}
\usepackage{tabularx}
\usepackage{pifont}
\usepackage{standalone}
\usepackage{rotating, multirow}
\usepackage{makecell}
\usepackage{subcaption}
\usepackage{hyperref}
\usepackage{titlesec}

\makeatletter
\def\blfootnote{\xdef\@thefnmark{}\@footnotetext}
\makeatother

\renewcommand*{\arraystretch}{0.95}
\DeclareMathAlphabet{\mathcal}{OMS}{cmsy}{m}{n}

\ifnum\isReadyToSubmit=0
 \usepackage[margin=1in]{geometry}
   \usepackage{fancyhdr}
\fi

\include{preamble}

\begin{document}

\date{}

\title{
  \Large \bf Stealing Machine Learning Models via Prediction
APIs}

\author{
{\rm Florian Tram\`{e}r}\\
EPFL
\and
{\rm Fan Zhang}\\
Cornell University
\and
{\rm Ari Juels}\\
Cornell Tech, Jacobs Institute
\and
{\rm Michael K.\ Reiter}\\
UNC Chapel Hill
\and
{\rm Thomas Ristenpart}\\
Cornell Tech
} 

\maketitle


\blfootnote{This is an extended version of a paper that appeared at USENIX 
Security, 2016.}

\input{abstract}

\input{intro}

\input{background}

\input{threatmodel}

\input{withconf}

\input{dectrees}
\input{ftransforms}
\input{labelonly}

\input{countermeasures}

\input{relwork}
\input{conclusion}

{\footnotesize \bibliographystyle{acm}
\bibliography{biblio}}

\appendix
\input{model-details}
\input{datasets}

\input{trees-analysis}

\input{improper}

\end{document}

%% file: preamble.tex
\newtheorem{theorem}{Theorem}

\newtheorem{definition}[theorem]{Definition}

\newcommand{\abs}[1]{\left| #1 \right|}

\newenvironment{proof-sketch}{\noindent{\bf Sketch of Proof}\hspace*{1em}}{\qed\bigskip}
\newenvironment{proof-idea}{\noindent{\bf Proof Idea}\hspace*{1em}}{\qed\bigskip}
\newenvironment{proof-of-lemma}[1]{\noindent{\bf Proof of Lemma #1}\hspace*{1em}}{\qed\bigskip}
\newenvironment{proof-attempt}{\noindent{\bf Proof Attempt}\hspace*{1em}}{\qed\bigskip}

\newcommand{\Reals}{\mathbb{R}}
\newcommand{\Ints}{\mathbb{Z}}

\newcommand*\OK{\ding{51}}
\newcommand*\NOK{\ding{55}}

\DeclareMathOperator*{\argmax}{argmax}

\newcommand{\set}[1]{\left\{ #1 \right\}}

\newcommand{\xvec}{\mathbf{x}}
\newcommand{\yvec}{\mathbf{y}}
\newcommand{\wvec}{\mathbf{w}}
\newcommand{\betavec}{\boldsymbol{\beta}}
\newcommand{\alphavec}{\boldsymbol{\alpha}}
\newcommand{\fspace}{\mathcal{X}}
\newcommand{\ispace}{\mathcal{M}}
\newcommand{\msg}{M}
\newcommand{\fdim}{d}
\newcommand{\range}{\mathcal{Y}}
\newcommand{\tree}{T}
\newcommand{\id}{\texttt{id}}
\newcommand{\oracle}{\mathcal{O}}
\newcommand{\ext}{\textsf{ex}}

\newcommand{\algorithmiccontinue}{\textbf{continue}}
\newcommand{\Continue}{\State \algorithmiccontinue}

\newcommand{\secref}[1]{\mbox{Section~\ref{#1}}}
\newcommand{\apref}[1]{\mbox{Appendix~\ref{#1}}}

\newcommand{\figref}[1]{\mbox{Figure~\ref{#1}}}
\renewcommand{\algref}[1]{\mbox{Algorithm~\ref{#1}}}
\renewcommand{\eqref}[1]{\mbox{(\ref{#1})}}

\newcommand{\tabref}[1]{\mbox{Table~\ref{#1}}}

\newcommand{\bnm}{\begin{newmath}}
\newcommand{\enm}{\end{newmath}}

\newcommand{\bea}{\begin{eqnarray*}}
\newcommand{\eea}{\end{eqnarray*}}

\newcommand{\bne}{\begin{newequation}}
\newcommand{\ene}{\end{newequation}}

\newenvironment{newmath}{\begin{displaymath}%
\setlength{\abovedisplayskip}{4pt}%
\setlength{\belowdisplayskip}{4pt}%
\setlength{\abovedisplayshortskip}{6pt}%
\setlength{\belowdisplayshortskip}{6pt} }{\end{displaymath}}

\newenvironment{newequation}{\begin{equation}%
\setlength{\abovedisplayskip}{4pt}%
\setlength{\belowdisplayskip}{4pt}%
\setlength{\abovedisplayshortskip}{6pt}%
\setlength{\belowdisplayshortskip}{6pt} }{\end{equation}}

\newcounter{ctr}

\newenvironment{newitemize}{%
\begin{list}{\mbox{}\hspace{0pt}$\bullet$\hfill}{\labelwidth=10pt%
\labelsep=0pt \leftmargin=10pt \topsep=1pt%
\setlength{\listparindent}{\saveparindent}%
\setlength{\parsep}{\saveparskip}%
\setlength{\itemsep}{1pt} }}{\end{list}}

\newenvironment{newenum}{%
\begin{list}{{\rm (\arabic{ctr})}\hfill}{\usecounter{ctr} \labelwidth=17pt%
\labelsep=0pt \leftmargin=22pt \topsep=1pt%
\setlength{\listparindent}{\saveparindent}%
\setlength{\parsep}{\saveparskip}%
\setlength{\itemsep}{1pt} }}{\end{list}}

\newlength{\saveparindent}
\newlength{\saveparskip}
\pgfplotscreateplotcyclelist{list}{%
{color=red,solid,mark=x},
{color=blue,dashed,every mark/.append style={solid},mark=diamond},
{color=teal,dotted,every mark/.append style={solid},mark=square},
{color=brown,dashdotted,every mark/.append style={solid},mark=o},
{color=black,dashdotdotted,every mark/.append style={solid},mark=triangle},
{color=orange,solid,every mark/.append style={solid},mark=*}}

\pgfplotsset{
  every axis plot/.append style={line width=0.8pt},
  every axis plot post/.append style={
    every mark/.append style={line width=0.8pt}
  }
}

\tikzset{
  treenode/.style = {shape=rectangle, rounded corners,
                     draw, black, align=center, inner sep=2pt,
                     top color=white, bottom color=blue!20},
  inner/.style      = {treenode, thin},
  leaf/.style    = {circle,black, thin, draw,inner sep=0.5pt},
  leafShade/.style    = {circle,OliveGreen,thick,draw,inner sep=0.5pt, text=black}
}

{\makeatletter
 \gdef\xxxmark{%
   \expandafter\ifx\csname @mpargs\endcsname\relax 
     \expandafter\ifx\csname @captype\endcsname\relax 
       \marginpar{\textcolor{red}{xxx~}}
     \else
       \textcolor{red}{xxx~}
     \fi
   \else
     \textcolor{red}{xxx~}
   \fi}
 \gdef\xxx{\@ifnextchar[\xxx@lab\xxx@nolab}
 \long\gdef\xxx@lab[#1]#2{{\bf [\xxxmark \textcolor{red}{#2} ---{\sc #1}]}}
 \long\gdef\xxx@nolab#1{{\bf [\xxxmark \textcolor{red}{#1}]}}
}

\def\authnotes{1}

\newcounter{notectr}[section]
\newcommand{\thenote}{\thesubsection.\arabic{notectr}\refstepcounter{notectr}}

\newcommand{\note}[2]{$\ll$#1~\thenote: #2$\gg$}
\newcommand{\tnote}[1]{\ifnum\authnotes=1 \textcolor{blue}{\note{TomR}{#1}}\fi}

\newcommand{\anote}[1]{\ifnum\authnotes=1 \textcolor{red}{\note{AJ}{#1}}\fi}
\newcommand{\fnote}[1]{\ifnum\authnotes=1 \textcolor{brown}{\note{FT}{#1}}\fi}
\newcommand{\znote}[1]{\ifnum\authnotes=1 \textcolor{purple}{\note{Fan}{#1}}\fi}

\newcommand{\outspace}{\mathcal{Y}}

\newcommand{\numclasses}{c}
\newcommand{\thh}{^\textrm{th}}
\newcommand{\R}{\mathbb{R}}

\newcommand{\fout}{\hat{f}}
\newcommand{\featext}{\phi}
\newcommand{\kernel}{K}
\newcommand{\kernelrbf}{\kernel_{\mathrm{rbf}}}
\newcommand{\kernelquad}{\kernel_{\mathrm{quad}}}
\newcommand{\sign}{\mathrm{sign}}

\newcommand{\splitfun}{\rho}

\newcommand{\trainalg}{\mathcal{T}}
\newcommand{\trainHide}{\mathcal{H}_\trainalg}

\newcommand{\rangedist}{d}

\newcommand{\setting}{\textsf{S}}
\newcommand{\testset}{D}
\newcommand{\unifset}{U}
\newcommand{\distclass}{\mathcal{P}}
\newcommand{\error}{R}
\newcommand{\errortest}{\error_{\mathrm{test}}}
\newcommand{\errortesttv}{\error_{\mathrm{test}}^{\mathrm{TV}}}
\newcommand{\errorunif}{\error_{\mathrm{unif}}}
\newcommand{\erroruniftv}{\error_{\mathrm{unif}}^{\mathrm{TV}}}
\newcommand{\traindist}{p}

\newcommand{\advA}{\mathcal{A}}

\newcolumntype{L}[1]{>{\raggedright\let\newline\\\arraybackslash\hspace{0pt}}p{#1}}
\newcolumntype{C}[1]{>{\centering\let\newline\\\arraybackslash\hspace{0pt}}p{#1}}
\newcolumntype{R}[1]{>{\raggedleft\let\newline\\\arraybackslash\hspace{0pt}}p{#1}}

\newcommand*\rotcol{\rotatebox{90}}

\newcommand{\ie}{{i.e.,~}}
\newcommand{\eg}{{e.g.,~}}

%% file: abstract.tex
\begin{abstract}

Machine learning (ML) models may be deemed confidential due to their sensitive training data, commercial value, or use in security applications. Increasingly often, confidential ML models are being deployed with publicly accessible query interfaces. ML-as-a-service (``predictive analytics'') systems are an example: Some allow users to train models on potentially sensitive data and charge others for access on a pay-per-query basis. 

The tension between model confidentiality and public access motivates our investigation of \emph{model extraction attacks}. In such attacks, an adversary with black-box access, but no prior knowledge of an ML model's parameters or training data, aims to duplicate the functionality of (\ie ``steal'') the model. Unlike in classical learning theory settings, ML-as-a-service offerings may accept partial feature vectors as inputs and include confidence values with predictions. Given these practices, we show simple, efficient attacks that extract target ML models with near-perfect fidelity for popular model classes including logistic regression, neural networks, and decision trees. We demonstrate these attacks against the online services of BigML and Amazon Machine Learning.
We further show that the natural countermeasure of omitting confidence
values from model outputs still admits potentially harmful
model extraction attacks.  Our results highlight the need for careful
ML model deployment and new model extraction countermeasures.
\end{abstract}

%% file: intro.tex
\section{Introduction}
\label{sec:intro}

Machine learning (ML) aims to provide automated extraction of insights from data by means of a predictive model. A predictive model is a function that maps feature vectors to a categorical or real-valued output. In a supervised setting, a previously gathered data set consisting of possibly confidential feature-vector inputs (e.g., digitized health records) with corresponding output class labels (e.g., a diagnosis) serves to train a predictive model that can generate labels on future inputs. Popular models include support vector machines (SVMs), logistic regressions, neural networks, and decision trees. 

ML algorithms' success in the lab and in practice has led to an explosion in demand. Open-source frameworks such as PredictionIO and cloud-based services offered by Amazon, Google, Microsoft,
BigML, and others have arisen to broaden and simplify ML model deployment. 

Cloud-based ML services often allow model owners to charge others for queries to their commercially valuable models. 
This pay-per-query deployment option exemplifies an increasingly common tension: The query interface of an ML model may be widely accessible, yet the model itself and the data on which it was trained may be proprietary and confidential. Models may also be privacy-sensitive because they leak information about
training data~\cite{fredrikson14,fredrikson15,ateniese2015hacking}. For security applications
such as spam or fraud
detection~\cite{lowd2005adversarial,biggio2013evasion,pdfrate14,
huang2011adversarial}, an ML model's confidentiality is critical to its utility: An adversary that can learn the model can also often evade detection~\cite{lowd2005adversarial,ateniese2015hacking}. 

In this paper we explore \emph{model extraction attacks}, which exploit the tension between query access and confidentiality in ML models. We consider an adversary that can query an ML model (a.k.a.~a prediction API) to obtain predictions on input feature vectors. The model may be viewed as a black box. The adversary may or may not know the model type (logistic regression, decision tree, etc.) or the distribution over the data used to train the model. The adversary's goal is to extract an equivalent or near-equivalent ML model, i.e.,
one that achieves (close to) 100\% agreement on an input space of interest. 

We demonstrate successful model extraction attacks against a wide variety of ML model types,
including decision trees, logistic regressions, SVMs, and deep neural networks,
and against production ML-as-a-service (MLaaS) providers, including Amazon  and BigML.\footnote{We simulated victims by training models in our
own accounts. We have
disclosed our results to affected services in February 2016.} In nearly all cases, our attacks yield models that are functionally very close to the target. 
In some cases, our attacks extract the exact parameters of the target
(e.g., the coefficients of a linear classifier or the paths of a decision tree). For some targets employing a model type, parameters or features unknown to the attacker, we additionally show a successful preliminary attack step involving reverse-engineering these model characteristics.

Our most successful attacks rely on the information-rich outputs returned by the ML prediction
APIs of all cloud-based services we investigated. Those of Google, Amazon, Microsoft, and
BigML all return {\em high-precision confidence values in addition to class labels}. They also respond to partial queries lacking one or more features. Our setting thus differs from traditional learning-theory settings
~\cite{valiant84,lowd2005adversarial, kushilevitz1993learning,jackson1994efficient,
bellare1992technique,angluin1988queries,
bshouty1995exact,benedek1991learnability} that assume
only {\em membership queries}, outputs consisting of a class label only. 
For example, for logistic regression, the confidence value is a simple log-linear function $1/(1+e^{-(\wvec \cdot \xvec + \beta)})$ of the $d$-dimensional
input vector $\xvec$. By querying $d+1$ random $d$-dimensional inputs, an attacker can with high probability solve for the unknown $d+1$ parameters~$\wvec$ and $\beta$ defining the 
model. 
We emphasize that while this model extraction attack is simple and non-adaptive, it affects all of the ML services we have investigated. 

Such equation-solving attacks extend to multiclass logistic regressions and neural networks, but do not work for decision trees, a popular model
choice. (BigML, for example, initially offered only decision trees.) For decision trees, a confidence value  reflects the number of training data points labeled correctly on an input's path in the tree; simple equation-solving  is thus inapplicable. We show how confidence values can nonetheless be exploited as pseudo-identifiers for paths in the tree, facilitating discovery of the tree's structure. We demonstrate successful model extraction attacks that use adaptive, iterative search algorithms to discover paths in a tree. 

\begin{table}
\center
\footnotesize
\begin{tabularx}{\columnwidth}{@{} @{\extracolsep{-5.5pt}} l b{2.35cm} l r >{\raggedleft\arraybackslash}X @{}}
\textbf{Service} & \textbf{Model Type} & \textbf{Data set} & \textbf{Queries} & \textbf{Time (s)}\\
\toprule
\multirow{2}{*}{Amazon} & Logistic Regression & Digits & $650$ & $70$ \\
& Logistic Regression
& Adult & $1{,}485$ & $149$ \\
\midrule
\multirow{2}{*}{BigML} & Decision Tree & German Credit & $1{,}150$ & $631$\\
& Decision Tree & Steak Survey & $4{,}013$ & $2{,}088$\\
\bottomrule
\end{tabularx}
\vspace{-10pt}
\caption{\footnotesize{\textbf{Results of model extraction attacks on ML services.} For each target model, we report the number of prediction queries made to the ML API in an attack that extracts a 100\% equivalent model. The attack time is primarily influenced by the service's prediction latency ($\approx 100\text{ms}/\text{query}$ for Amazon and $\approx 500\text{ms}/\text{query}$ for BigML).}}
\label{tab:result-summary}
\end{table}

We experimentally evaluate our attacks by training models on an array of public
data sets suitable as stand-ins for proprietary ones. We validate the
attacks locally using standard ML libraries, and then present case studies 
on BigML and Amazon. For both services, we show computationally fast attacks that use a small number of queries to extract models matching the targets on 100\% of tested inputs. 
See \tabref{tab:result-summary} for a quantitative summary. 

Having demonstrated the broad applicability of model extraction attacks to existing services, we consider the most obvious potential countermeasure ML services might adopt: Omission of confidence values, i.e., output of class labels only. This approach would place model extraction back in the membership query setting of prior work in learning theory~\cite{valiant84,lowd2005adversarial,
angluin1988queries,benedek1991learnability}. We demonstrate a generalization of an adaptive algorithm by Lowd and Meek~\cite{lowd2005adversarial} from binary linear classifiers to more complex model types, and also propose an attack inspired by the agnostic learning algorithm of Cohn et al.~\cite{cohn1994improving}. Our new attacks extract models matching targets on $>$99\% of the input space for a variety of model classes, but need up to $100\times$ more queries than equation-solving attacks (specifically for multiclass linear regression and neural networks). While less effective than equation-solving, these attacks remain attractive for certain types of adversary. We thus discuss further ideas for
countermeasures.

In summary, we explore model extraction attacks, a practical kind of learning
task that, in particular, affects emerging cloud-based ML services being built by Amazon,
Google, Microsoft, BigML, and others. We show:
\begin{newitemize}
\item \emph{Simple equation-solving model extraction attacks} that use non-adaptive, random queries to solve for the
parameters of a target model. These attacks affect a
wide variety of ML models that output confidence values. We show their success against Amazon's service (using our own
models as stand-ins for victims'), and also report successful reverse-engineering of the (only partially documented) model type employed by Amazon.
\item \emph{A new path-finding algorithm for extracting decision trees}
that abuses confidence values as quasi-identifiers for paths. To our knowledge, this is the first example of practical ``exact'' decision tree
learning. We demonstrate the attack's efficacy via experiments on BigML.
\item \emph{Model extraction attacks against models that output only class labels}, the obvious countermeasure against extraction attacks that rely on confidence values. We show slower, but still potentially dangerous, attacks in this setting that build on prior work in learning theory.
\end{newitemize}
We additionally make a number of observations about the implications
of extraction. For example, attacks against Amazon's system 
indirectly leak various summary statistics about a private training
set, while extraction against kernel logistic regression models~\cite{zhu01}
recovers significant information about individual training data points. 

The source code for our attacks is available online at \url{https://github.com/ftramer/Steal-ML}.

%% file: background.tex
\section{Background}
\label{sec:prelims}

For our purposes, a ML model is a function $f\colon\fspace\rightarrow\outspace$.
An input is a $\fdim$-dimensional vector in the feature space 
$\fspace = \fspace_1\times \fspace_2 \times \dots \times \fspace_\fdim$. Outputs lie in the range
$\outspace$.

We distinguish between categorical features, which assume one of a
finite set of values (whose set size is the arity of the feature), 
and continuous features, which assume a value in a bounded subset of the
real numbers. Without loss of generality, for a categorical feature of arity $k$, we let $\fspace_i = \Ints_k$.
For a continuous feature taking values between bounds $a$ and $b$, we let
$\fspace_i = [a,b] \subset \Reals$.  
Inputs to a model may be pre-processed to perform feature extraction. In this case, inputs come from a space $\ispace$, and feature extraction involves application of a function $\ext\colon\ispace\rightarrow\fspace$ that maps inputs into a feature space. Model application then proceeds by composition in the natural way, taking the form $f(\ext(\msg))$.
Generally, feature extraction is many-to-one. For example, $\msg$ 
may be a piece of English language text and the extracted features counts of individual words (so-called ``bag-of-words'' feature extraction). Other examples are input scaling and one-hot-encoding of categorical features.

We focus primarily on classification settings in which $f$
predicts a nominal variable ranging over
a set of classes.  Given  $\numclasses$ classes, we use as class labels the set
$\Ints_\numclasses$.  If $\outspace = \Ints_\numclasses$, the model returns only the predicted class label. In some
applications, however, additional information is often helpful, in the form of
real-valued measures of confidence on the labels output by the model; these measures are called \emph{confidence values}.  The output space is then $\outspace =
[0,1]^{\numclasses}$. For a given $\xvec \in \fspace$ and $i \in \Ints_\numclasses$, we denote by $f_i(\xvec)$ the
$i\thh$ component of $f(\xvec) \in \outspace$.  The value
$f_i(\xvec)$ is a model-assigned probability that $\xvec$ has associated class label $i$.  The model's predicted class is defined by the value $\argmax_{i}\; f_i(\xvec)$, \ie the most probable label. 

We associate with $\outspace$ a distance measure
$\rangedist_\outspace$. We drop the subscript $\outspace$ when it is clear from
context. For
$\outspace = \Ints_\numclasses$ we use 0-1 distance, meaning $\rangedist(y,y') =
0$ if $y = y'$ and $\rangedist(y,y') = 1$ otherwise. For $\outspace = [0,1]^{\numclasses}$,
we use the 0-1 distance when comparing predicted classes; when comparing class probabilities directly, we instead use the total
variation distance, given by $\rangedist(\yvec,\yvec') = \frac12 \sum |\yvec[i]
- \yvec'[i]|$. In the rest of
this paper, unless explicitly specified otherwise, $d_\outspace$ refers to the 0-1 distance over class labels.


\paragraph{Training algorithms.}
We consider models obtained via supervised learning. These models are generated by a training algorithm $\trainalg$ that takes as input a training set $\{(\xvec_i,y_i)\}_i$, where  $(\xvec_i,y_i) \in
\fspace\times\outspace$ is an input with an associated (presumptively correct) class label. The output of $\trainalg$ is a model $f$
defined by a set of \emph{parameters}, which are model-specific, and \emph{hyper-parameters}, which specify the type of models $\trainalg$ generates.
Hyper-parameters may be viewed as distinguished parameters, often taken from a small number of
standard values; for example, the kernel-type used in an SVM, of which only a small set are used in practice, may be seen as a hyper-parameter. 




%% file: threatmodel.tex
\section{Model Extraction Attacks}
\label{sec:threatmodel}

\begin{figure}[t]
\center
\includegraphics[width=2.5in]{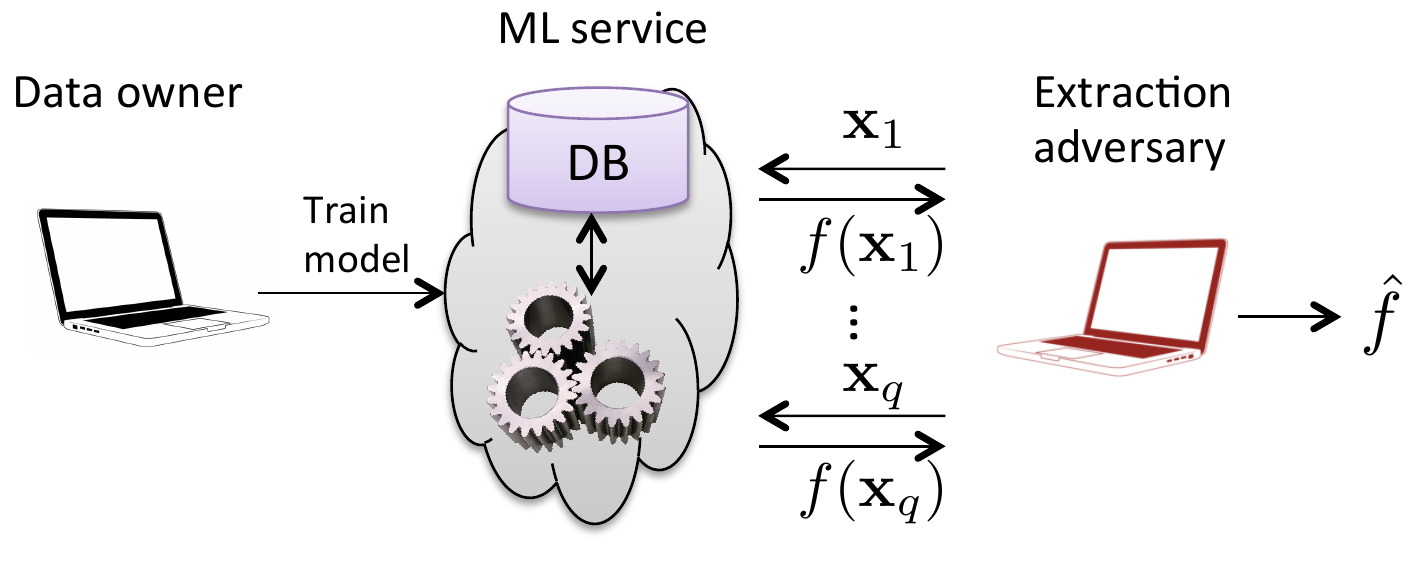}
\vspace{-12pt}
\caption{\footnotesize{\textbf{Diagram of ML model extraction attacks.} A data owner has a model
$f$ trained on its data and allows others to make prediction queries.
An adversary uses $q$ prediction queries to extract an $\fout \approx f$.
}}
\label{fig:diagram}
\end{figure}

An ML model extraction attack arises when an adversary obtains
black-box access to some \emph{target model} $f$ and attempts to learn a model $\fout$
that closely approximates, or even matches, $f$ (see \figref{fig:diagram}). 

As mentioned previously, the restricted case in which $f$ outputs class labels only, matches the membership query setting considered in learning theory, \eg
PAC learning~\cite{valiant84} and other previous works~\cite{lowd2005adversarial, kushilevitz1993learning,jackson1994efficient,
bellare1992technique,angluin1988queries, bshouty1995exact,benedek1991learnability}. Learning theory algorithms have seen only limited study in practice, e.g., in~\cite{lowd2005adversarial}, and our investigation may be viewed as a practice-oriented exploration of this branch of research. Our initial focus, however, is on a different setting common in today's MLaaS services, which we now explain in detail. Models trained by these services emit data-rich outputs that often include confidence values, and in which partial feature vectors may be considered valid inputs. As we show later, this setting greatly advantages adversaries.

\paragraph{Machine learning services.} A number of companies have launched or
are planning to launch cloud-based ML services. A common denominator is the
ability of users to upload data sets, have the provider
run training algorithms on the data, and make the resulting models generally available
for prediction queries. Simple-to-use Web APIs handle the entire interaction.
This service model lets users capitalize on their data without having to
set up their own large-scale ML infrastructure.
Details vary greatly across services. We summarize a number of them in
\tabref{tab:api-summary} and now explain some of the salient features. 

\begin{table}
\center\footnotesize
\def\arraystretch{0.9}

\renewcommand\theadalign{lb}
\settowidth{\rotheadsize}{\textbf{Confidence}}
\setlength{\tabcolsep}{4pt}
\begin{tabularx}{\columnwidth}{@{} X @{\hskip 5pt} c c c @{\hskip 0.2in} c c c c @{}}
\thead{\textbf{Service}} & \rothead{\textbf{White-box}} & \rothead{\textbf{Monetize}} &
\rothead{\textbf{Confidence Scores}} & \rothead{\textbf{Logistic Regression}} & \rothead{\textbf{SVM}} & \rothead{\textbf{Neural Network}} & \rothead{\textbf{Decision Tree}}\vspace{-3pt}\\
\toprule
Amazon~\cite{aws} & \NOK & \NOK & \OK & \OK & \NOK & \NOK & \NOK \\
Microsoft~\cite{azureml} & \NOK & \NOK & \OK & \OK & \OK & \OK & \OK \\
BigML~\cite{bigml} & \OK & \OK & \OK & \OK & \NOK & \NOK & \OK \\
PredictionIO~\cite{predictionio} & \OK & \NOK & \NOK & \OK & \OK & \NOK & \OK \\
Google~\cite{googlepred} & \NOK & \OK & \OK & \OK & \OK & \OK & \OK \\
\bottomrule
\end{tabularx}
\vspace{-9pt}
\caption{\footnotesize{\textbf{Particularities of major MLaaS providers.} `White-box' refers to the ability to download and use a trained model locally, and `Monetize' means that a user may charge other users for black-box access to her models. 
Model support for each service is obtained from available documentation. The models listed for Google's API are a projection based on the announced support of models in standard PMML format~\cite{googlepred}. Details on ML models are given in \apref{sec:model-details}.
}}
\label{tab:api-summary}
\end{table}

A model is \emph{white-box} if a user may download a representation
suitable for local use. It is \emph{black-box} if accessible only via a prediction query interface. 
Amazon and Google, for example, provide black-box-only services. Google does not even specify what training algorithm their service uses, while Amazon provides only partial documentation for its feature extraction $\ext$ (see \secref{sec:real-world}).
Some services allow users to monetize trained models by charging others for prediction queries. 

To use these services, a user uploads a data set and optionally applies some data pre-processing (\eg field removal or handling of missing values). She then trains a model by either choosing one of many supported model classes (as in BigML, Microsoft, and PredictionIO) or having the service choose an appropriate model class (as in Amazon and Google). Two services have also announced upcoming support for users to upload their own trained models (Google) and their own custom learning algorithms (PredictionIO). 
When training a model, users may tune various parameters of the model or training-algorithm (\eg regularizers,  tree size, learning rates) and control feature-extraction and transformation methods.

For black-box models, the service provides users with information needed to create and interpret predictions, such as the list of input features and their types. Some services also supply the model class, chosen training parameters, and training data statistics (\eg BigML gives the range, mean, and standard deviation of each feature).

To get a prediction from a model, a user sends one or more input queries. The services we reviewed accept both synchronous requests and asynchronous `batch' requests for multiple predictions. We further found varying degrees of support for `incomplete' queries, in which some input features are left unspecified~\cite{saar07}. We will show that exploiting incomplete queries can drastically improve the success of some of our attacks. Apart from PredictionIO, all of the services we examined respond to prediction queries with not only class labels, but a variety of additional information, including \emph{confidence scores} (typically class probabilities) for the predicted outputs.

Google and BigML allow model owners to monetize their models by charging other users for predictions. Google sets a minimum price of $\$0.50$ per $1{,}000$ queries. On BigML, $1{,}000$ queries consume at least $100$ \emph{credits}, costing $\$0.10$--$\$5$, depending on the user's subscription.

\paragraph{Attack scenarios.} 
We now describe possible motivations for adversaries to perform model extraction attacks. We then present a
more detailed threat model informed by characteristics of the aforementioned ML services.

\emph{Avoiding query charges.} Successful monetization of
prediction queries by the owner of an ML model $f$ requires confidentiality of $f$. 
A malicious user may seek to launch what we call a \emph{cross-user} model extraction attack, stealing
$f$ for subsequent free use. 
More subtly, in black-box-only settings (\eg Google and Amazon),
a service's business model may involve
amortizing up-front training costs by charging users for future predictions. A model extraction attack will undermine the provider's business
model if a malicious user pays less for training and extracting than for paying per-query charges. 


\emph{Violating training-data privacy.} Model extraction could, in turn, leak information about sensitive training data. Prior
attacks such as model inversion~\cite{fredrikson14,fredrikson15,ateniese2015hacking} have shown that
access to a model can be abused to infer information about training set points. Many of these attacks work better in white-box settings;
model extraction may thus be a stepping stone to such privacy-abusing
attacks. Looking ahead, we will see that in some cases, significant information
about training data is leaked trivially by successful model extraction, because
the model itself directly incorporates training set points.

\emph{Stepping stone to evasion.} In settings
where an ML model serves to detect adversarial behavior, such as identification of
spam, malware classification, and network anomaly detection, model extraction can facilitate \emph{evasion attacks}. An adversary may use knowledge of the ML model to avoid detection by it~\cite{ateniese2015hacking,huang2011adversarial,lowd2005adversarial,biggio2013evasion,pdfrate14}. 


In all of these settings, there is an inherent assumption of secrecy of the ML model in use. We show that this assumption is broken for all ML APIs that we investigate.

\paragraph{Threat model in detail.} Two distinct adversarial models arise in practice. An adversary may be able to make \emph{direct} queries, providing an arbitrary input $\xvec$ to a model $f$ and obtaining the output $f(\xvec)$. Or the adversary may be able to make only \emph{indirect} 
queries, \ie queries on points in input space $\ispace$ yielding outputs $f(\ext(\msg))$. The feature extraction mechanism $\ext$ may be unknown to the adversary. In~\secref{sec:real-world}, we show how ML APIs can further be exploited to ``learn''  feature extraction mechanisms. Both direct and indirect 
access to $f$ arise in ML services. (Direct query interfaces arise when clients are expected to perform feature extraction locally.) In either case, the output value can be a class label, a
confidence value vector, or some data structure revealing various
levels of information, depending on the exposed API.


We model the adversary, denoted by~$\advA$,  as a
randomized algorithm. The adversary's goal is to
use as few queries as possible to $f$ in order to efficiently compute an
approximation $\fout$ that closely matches $f$.
We formalize ``closely matching'' using two different error measures: 

\begin{newitemize}
\item \emph{Test error $\errortest$}: This is the average error over
a test set $\testset$, given by 
$\errortest(f,\fout) =$ $\sum_{(\xvec,y)\in\testset}
\rangedist(f(\xvec),\fout(\xvec)) / |\testset|$. 
A low test error implies
that $\fout$ matches $f$ well for inputs distributed
like the training data samples.~\footnote{Note that for some $\testset$, it is possible that 
$\fout$ predicts true labels better than $f$, yet $\errortest(f,\fout)$ is large, because $\fout$ does not closely match $f$.}

%
\item \emph{Uniform error $\errorunif$}:  
For a set $\unifset$ of vectors uniformly chosen in $\fspace$, let $\errorunif(f,\fout)
= \sum_{\xvec\in\unifset} \rangedist(f(\xvec),\fout(\xvec)) / |\unifset|$. 
Thus $\errorunif$ estimates the fraction of the full feature space on which 
$f$ and $\fout$ disagree. (In our experiments, we found $|\unifset| = 10,000$ was sufficiently large to obtain stable error estimates for the models we analyzed.) 
\end{newitemize}

We define the extraction \emph{accuracy} under test and uniform error as $1 - \errortest(f,\fout)$ and $1-\errorunif(f,\fout)$. Here we implicitly refer to accuracy under 0-1 distance. When assessing how close the class probabilities output by $\fout$ are to those of $f$ (with the total-variation distance) we use the notations $\errortesttv(f,\fout)$ and $\erroruniftv(f,\fout)$.

An adversary may know any of a number of pieces of information about a target $f$: What training algorithm $\trainalg$
generated $f$, the hyper-parameters used with $\trainalg$, 
the feature extraction function $\ext$, etc. We will
investigate a variety of settings in this work corresponding to different APIs seen in practice. We assume that $\advA$ has no more information about a model's training data,
than what is provided by an ML API (\eg summary statistics).
For simplicity, we focus
on \emph{proper} model extraction: If $\advA$ believes that $f$ belongs to some model class, then $\advA$'s goal is to extract a model $\fout$ from the
\emph{same} class. We discuss some intuition in favor of proper extraction in \apref{sec:improper}, 
and leave a broader treatment of \emph{improper} extraction strategies as an interesting open problem.

%% file: withconf.tex
\section{Extraction with Confidence Values}
\label{sec:withconf}

We begin our study of extraction attacks by focusing on prediction APIs that return confidence values. As per \secref{sec:prelims}, the output 
of a query to $f$ thus falls in a range $[0,1]^\numclasses$ where
$\numclasses$ is the number of classes. 
To motivate this, we recall that most ML APIs reveal confidence values for
models that support them (see \tabref{tab:api-summary}). This includes logistic regressions (LR), neural networks, and decision trees, defined formally in \apref{sec:model-details}.
We first introduce a generic \emph{equation-solving}
attack that applies to all logistic models (LR and neural networks). In \secref{sec:dectrees}, we present two novel \emph{path-finding} attacks on decision trees.

\subsection{Equation-Solving Attacks}
\label{sec:eqsolve}

Many ML models we consider directly compute class probabilities as a continuous function of the input $\xvec$ and real-valued model parameters. In this case, an API that reveals these class probabilities provides an adversary $\advA$ with samples $(\xvec, f(\xvec))$ that can be viewed as equations in the unknown model parameters. For a large class of models, these equation systems can be efficiently solved, thus recovering $f$ (or some good approximation of it).

Our approach for evaluating attacks will primarily be experimental. We
use a suite of synthetic or publicly available data sets to serve as
stand-ins for proprietary data that might be the target of an extraction attack.
\tabref{tab:benchmark} displays the data sets used in this section, which we obtained from various sources: the synthetic ones we generated;
the others are taken from public surveys (\emph{Steak Survey}~\cite{538-survey} and \emph{GSS Survey}~\cite{gss}), from \texttt{scikit}~\cite{scikit} (\emph{Digits}) or from the UCI ML library~\cite{uci}.
Mre details about these data sets are in \apref{sec:datasets}.

\begin{table}[t]
\center
\def\arraystretch{0.8}
\footnotesize
\begin{tabularx}{\columnwidth}{@{} l c *{3}{>{\raggedleft\arraybackslash}X @{}}}
\textbf{Data set} & \textbf{Synthetic} & \textbf{\# records} & \textbf{\# classes} & \textbf{\# features} \\
\toprule
Circles & Yes & $5{,}000$ & $2$ & $2$\\
Moons & Yes & $5{,}000$ & $2$ & $2$\\
Blobs & Yes & $5{,}000$ & $3$ & $2$\\
$5$-Class & Yes & $1{,}000$ & $5$ & $20$\\
\midrule
Adult (Income) & No & $48{,}842$ & $2$ & $108$\\
Adult (Race) & No & $48{,}842$ & $5$ & $105$\\
Iris & No & $150$ & $3$ & $4$\\
Steak Survey & No & $331$ & $5$ & $40$\\
GSS Survey & No & $16{,}127$ & $3$ & $101$\\
Digits & No & $1{,}797$ & $10$ & $64$\\
Breast Cancer & No & $683$ & $2$ & $10$ \\
Mushrooms & No & $8{,}124$ & $2$ & $112$ \\
Diabetes & No & $768$ & $2$ & $8$\\
\bottomrule
\end{tabularx}
\vspace{-10pt}
\caption{\scriptsize{\textbf{Data sets used for extraction attacks.} 
We train two models on the Adult data, with targets `Income' and `Race'.
SVMs and binary logistic regressions are trained on data sets with $2$ classes. Multiclass regressions and neural networks are trained on multiclass data sets. For decision trees, we use a set of public models shown in \tabref{tab:benchmark-trees}.}} 
\label{tab:benchmark}
\end{table}

Before training, we remove rows with missing
values, apply \emph{one-hot-encoding} to categorical features, and scale all
numeric features to the range $[-1, 1]$. We train our models
over a randomly chosen subset of $70\%$ of the data, and keep the rest for
evaluation (\ie to calculate $\errortest$). 
We discuss the impact of different
pre-processing and feature extraction steps in \secref{sec:real-world}, when we evaluate equation-solving attacks 
on production ML services.

\subsubsection{Binary logistic regression}
As a simple starting point, we consider the case of logistic regression (LR). 
A LR model performs binary classification $(\numclasses = 2)$,
by estimating the probability of a binary response, based on a number of independent 
features. LR is one of the most popular binary classifiers, 
due to its simplicity and efficiency. It is widely used in many 
scientific fields (\eg medical and social sciences) and is supported by all the ML
services we reviewed.

Formally, a LR model is defined by parameters $\wvec \in \R^d$, $\beta \in \R$, and outputs a probability
$f_1(\xvec) = \sigma(\wvec\cdot\xvec+\beta)$, where $\sigma(t)= 1/(1+e^{-t})$.
LR is a linear classifier: it defines a \emph{hyperplane} in
the feature space $\fspace$ (defined by $\wvec\cdot\xvec+\beta=0$), that separates the two classes.

Given an oracle sample $(\xvec, f(\xvec))$, we get a \emph{linear} equation $\wvec \cdot \xvec + \beta =
\sigma^{-1}(f_1(\xvec))$. Thus, $\fdim+1$ samples
are both necessary and sufficient (if the queried $\xvec$ are linearly
independent) to recover $\wvec$ and $\beta$.
Note that the required samples are chosen non-adaptively, and can thus be
obtained from a single batch request to the ML service.

We stress that while this extraction attack is rather straightforward, it
 directly applies, with possibly devastating consequences, to all cloud-based ML
services we considered. As an example, recall that some services (\eg BigML and
Google) let model owners monetize black-box access to their models. Any user who
wishes to make more than $\fdim+1$ queries to a model would then minimize the
prediction cost by first running a cross-user model extraction attack, and then using the
extracted model for personal use, free of charge.
As mentioned in \secref{sec:threatmodel}, attackers with a final goal of 
model-inversion or evasion may also have incentives to first extract the model.
Moreover, for services with black-box-only access (\eg Amazon or Google), a user may abuse the service's resources to
train a model over a large data set $\testset$ (\ie $|\testset| \gg \fdim$), and extract it after only $\fdim+1$ predictions. Crucially, the extraction cost is independent of $|\testset|$. This could undermine a service's business model, should prediction fees be used to amortize the high cost of training.

For each binary data set shown in \tabref{tab:benchmark}, we train a LR model 
and extract it given $\fdim+1$ predictions. In all cases, we achieve $\errortest=\errorunif=0$. If we
compare the probabilities output by $f$ and $\fout$, $\errortesttv$ and $\erroruniftv$ are lower
than $10^{-9}$. For these models, the attack requires 
only $41$ queries on average, and $113$ at most.
On Google's platform for example, an extraction attack
would cost less than $\$0.10$, and subvert any further model monetization.

\subsubsection{Multiclass LRs and Multilayer Perceptrons}
\label{sec:withconf:multiclass}

We now show that such equation-solving attacks broadly extend to all model classes with a `logistic' layer, including multiclass $(\numclasses > 2)$ LR and deeper neural networks. We define these models formally in \apref{sec:model-details}. 

A multiclass logistic regression (MLR) combines $\numclasses$ binary models, each with parameters $\wvec_i, \beta_i$, to form a multiclass model. MLRs are available in all ML services we reviewed.
We consider two types of MLR models: softmax and one-vs-rest (OvR), 
that differ in how the $\numclasses$ binary models are trained and combined: A softmax model fits a joint multinomial distribution to all training samples, while a OvR model trains a separate binary LR for each class, and then normalizes the class probabilities.

A MLR model $f$ is defined by parameters $\wvec \in \R^{\numclasses \fdim}$, $\betavec \in \R^{\numclasses}$. Each sample $(\xvec, f(\xvec))$ gives $c$ equations in $\wvec$ and $\betavec$. The equation system is non-linear however, and has no analytic solution. For softmax models for instance, the equations take the form $e^{\wvec_i\cdot \xvec + \beta_i} / (
  \sum_{j=0}^{\numclasses-1} e^{\wvec_j\cdot \xvec + \beta_j}) = f_i(\xvec)$.
A common method for solving such a system is by minimizing an appropriate loss function, such as the logistic loss.
With a regularization term, the loss function is \emph{strongly convex}, and the optimization thus converges to a \emph{global minimum} (\ie a function $\fout$ that predicts the same probabilities as $f$ for all available samples). A similar optimization (over class labels rather than probabilities) is actually used for training logistic models. Any MLR implementation can thus easily be adapted for model extraction with equation-solving.


This approach naturally extends to deeper neural networks. We consider multilayer perceptrons (MLP), that first apply a non-linear transform to all inputs (the hidden layer), followed by a softmax regression in the transformed space. MLPs are becoming increasingly popular due to the continued success of deep learning methods; the advent of cloud-based ML services is likely to further boost their adoption.
For our attacks, MLPs and MLRs mainly differ in the number of unknowns in the system to solve. For perceptrons with one hidden layer, we have $\wvec \in \R^{\fdim h + h\numclasses}$, $\betavec \in \R^{h + \numclasses}$, where $h$ is the number of hidden nodes ($h=20$ in our experiments). Another difference is that the loss function for MLPs is not strongly convex. The optimization may thus converge to a local minimum, \ie a model $\fout$ that does not exactly match $f$'s behavior.

\begin{table}[t]
\footnotesize
\def\arraystretch{0.9}
\begin{tabularx}{\columnwidth}{@{} @{\extracolsep{-0.1cm}} l r r r r >{\raggedleft\arraybackslash}X @{}}
\textbf{Model} & \textbf{Unknowns} & \textbf{Queries} & $1-\errortest$ & $1-\errorunif$ & \textbf{Time (s)}\\[1pt]
\toprule
\multirow{2}{*}{Softmax} & \multirow{2}{*}{$530$} & 
    $265$ & $99.96\%$ & $99.75\%$ & $2.6$\\
& & $530$ & $100.00\%$ & $100.00\%$ & $3.1$\\
\midrule
\multirow{2}{*}{OvR} & \multirow{2}{*}{$530$} & 
    $265$ & $99.98\%$ & $99.98\%$ & $2.8$\\
& & $530$ & $100.00\%$ & $100.00\%$ & $3.5$\\
\midrule
\multirow{4}{*}{MLP} & \multirow{4}{*}{$2{,}225$} 
  & $1{,}112$ & $98.17\%$ & $94.32\%$ & $155$\\
& & $2{,}225$ & $98.68\%$ & $97.23\%$ & $168$\\
& & $4{,}450$ & $99.89\%$ & $99.82\%$ & $195$\\
& & $11{,}125$ & $99.96\%$ & $99.99\%$ & $89$\\
\bottomrule
\end{tabularx}
\vspace{-10pt}
\caption{\footnotesize{\textbf{Success of equation-solving attacks.} Models to extract were trained on the Adult data set with multiclass target `Race'. For each model, we report the number of unknown model parameters, the number of queries used, and the running time of the equation solver. The attack on the MLP with $11{,}125$ queries converged after $490$ epochs.}}
\label{tab:eq-solving}
\end{table}

To illustrate our attack's success, we train a softmax regression, a OvR regression and a MLP on the Adult data set with target `Race' ($\numclasses = 5$).
For the non-linear equation systems we obtain, we do not know a priori how many samples we need to find a solution (in contrast to linear systems where $\fdim + 1$ samples are necessary and sufficient). We thus explore various query budgets of the form $\alpha \cdot k$, where $k$ is the number of unknown model parameters, and $\alpha$ is a budget scaling factor.
For MLRs, we solve the equation system with BFGS~\cite{numericopti} in \texttt{scikit}~\cite{scikit}. For MLPs, we use \texttt{theano}~\cite{theano} to run stochastic gradient descent for $1{,}000$ epochs. Our experiments were performed on a commodity laptop (2-core Intel CPU @3.1GHz, 16GB RAM, no GPU acceleration).

\tabref{tab:eq-solving} shows the extraction success for each model, as we vary $\alpha$ from $0.5$ to at most $5$.
For MLR models (softmax and OvR), the attack is extremely efficient, requiring around one query per unknown parameter of $f$ (each query yields $c=5$ equations). 
For MLPs, the system to solve is more complex, with about $4$ times more unknowns. With a sufficiently over-determined system, we converge to a model $\fout$ that very closely approximates $f$. 
As for LR models, queries are chosen non-adaptively, so $\advA$ may submit a single `batch request' to the API.

We further evaluated our attacks over all multiclass data sets from \tabref{tab:benchmark}. For MLR models with $k=\numclasses\cdot (\fdim+1)$ parameters ($c$ is the number of classes), $k$ queries were sufficient to achieve perfect extraction ($\errortest = \errorunif = 0$, $\errortesttv$ and $\erroruniftv$ below $10^{-7}$). We use $260$ samples on average, and $650$ for the largest model (Digits). For MLPs with 20 hidden nodes, we achieved $>$99.9\% accuracy with $5{,}410$ samples on average and $11{,}125$ at most (Adult). With $54{,}100$ queries on average, we extracted a $\fout$ with 100\% accuracy over tested inputs.
As for binary LRs, we thus find that cross-user model extraction attacks for these model classes can be extremely efficient.


\subsubsection{Training Data Leakage for Kernel LR}
\label{sec:withconf:klr}

We now move to a less mainstream model class, \emph{kernel
logistic regression}~\cite{zhu01}, that illustrates
how extraction attacks can leak private training data, when a
model's outputs are directly computed as a function of that data.

Kernel methods are commonly used to efficiently extend support vector machines (SVM) to nonlinear classifiers~\cite{boser1992training}, but similar techniques can be applied to logistic regression~\cite{zhu01}. Compared to kernel SVMs, kernel logistic regressions (KLR) have the advantage of computing class probabilities, and of naturally extending to multiclass problems. Yet, KLRs have not reached the popularity of kernel SVMs or standard LRs, and are not provided by any MLaaS provider at the time. We note that KLRs could easily be constructed in any ML library that supports both kernel functions and LR models.

\begin{figure}

\newlength{\imsize}
\setlength{\imsize}{0.6cm}
\footnotesize
\centering
\begin{subfigure}[b]{0.45\columnwidth}

\begin{subfigure}[b]{\imsize}
	\centering
	\includegraphics[width=\imsize]{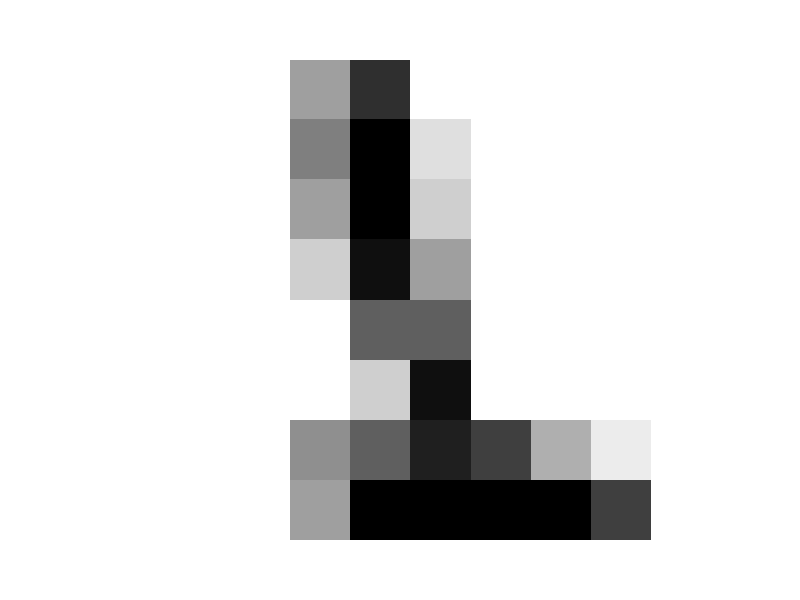}
\end{subfigure}%
\hfill
\begin{subfigure}[b]{\imsize}
	\centering
	\includegraphics[width=\imsize]{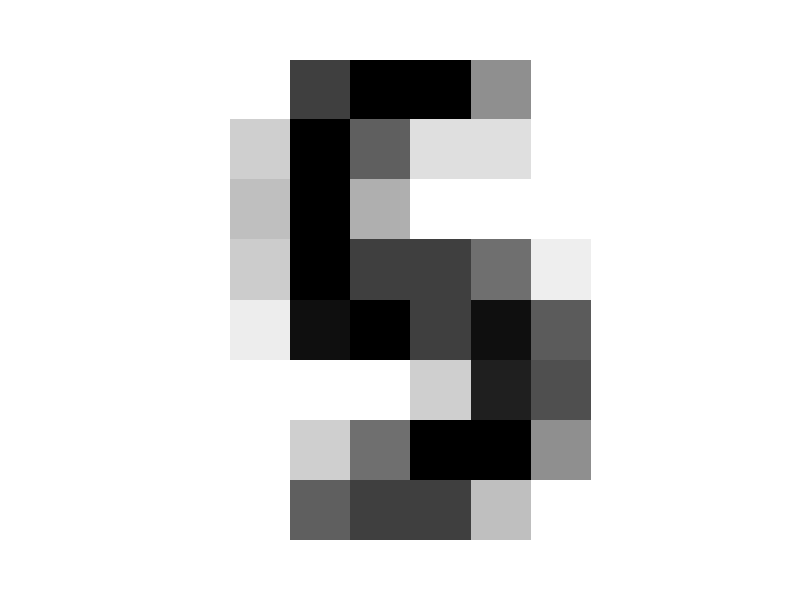}
\end{subfigure}%
\hfill
\begin{subfigure}[b]{\imsize}
	\centering
	\includegraphics[width=\imsize]{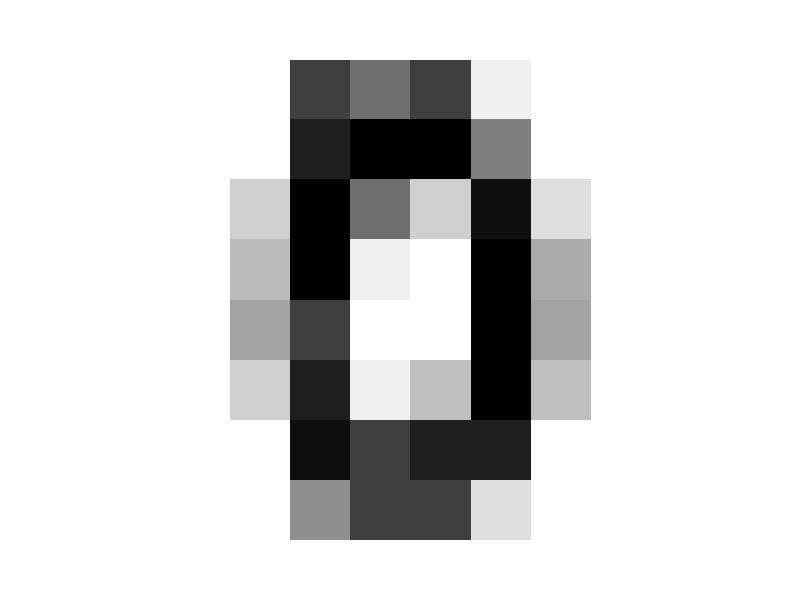}
\end{subfigure}%
\hfill
\begin{subfigure}[b]{\imsize}
	\centering
	\includegraphics[width=\imsize]{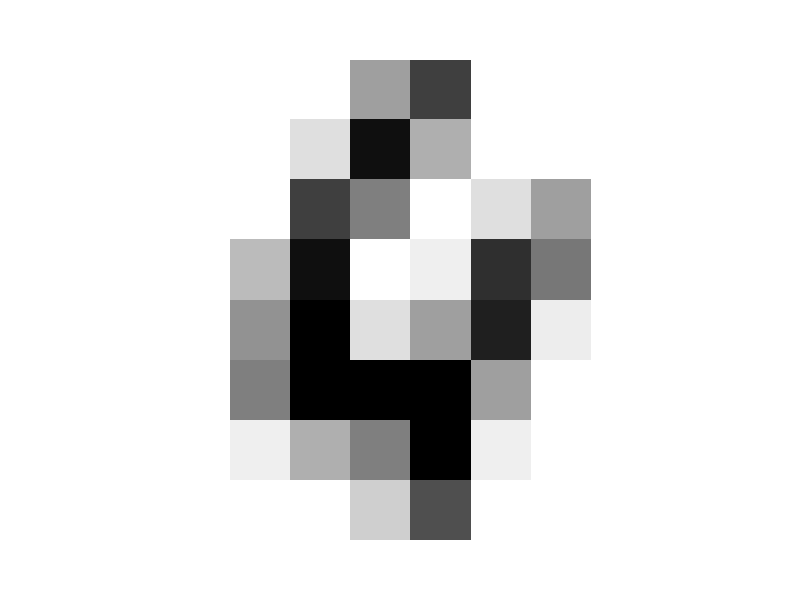}
\end{subfigure}%
\hfill
\begin{subfigure}[b]{\imsize}
	\centering
	\includegraphics[width=\imsize]{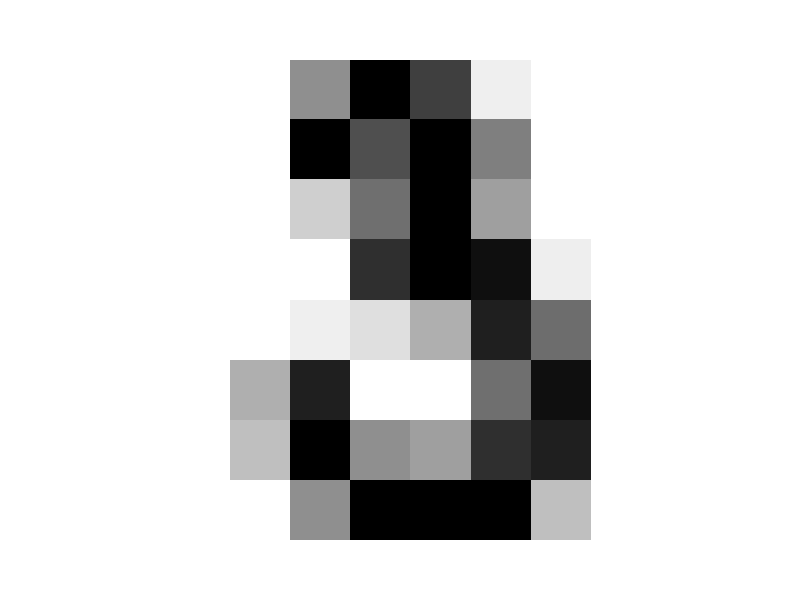}
\end{subfigure}%
\\
\begin{subfigure}[b]{\imsize}
	\centering
	\includegraphics[width=\imsize]{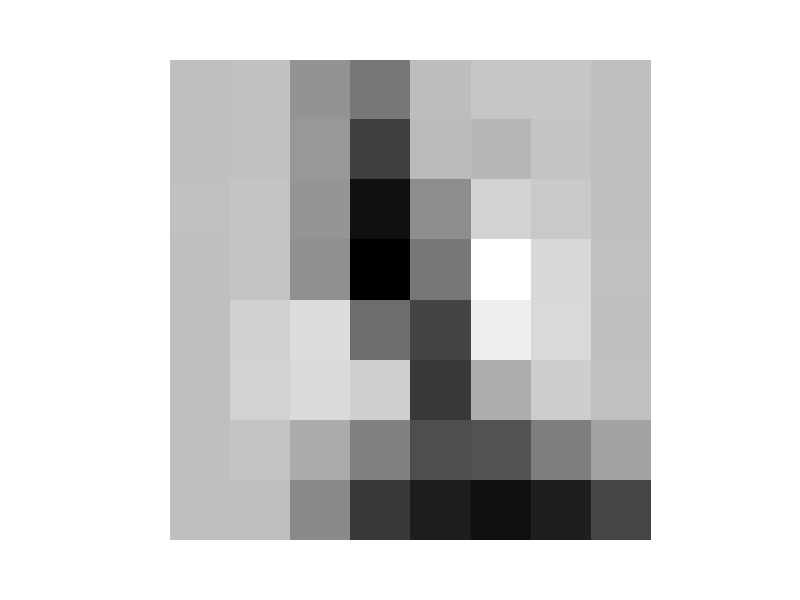}
\end{subfigure}%
\hfill
\begin{subfigure}[b]{\imsize}
	\centering
	\includegraphics[width=\imsize]{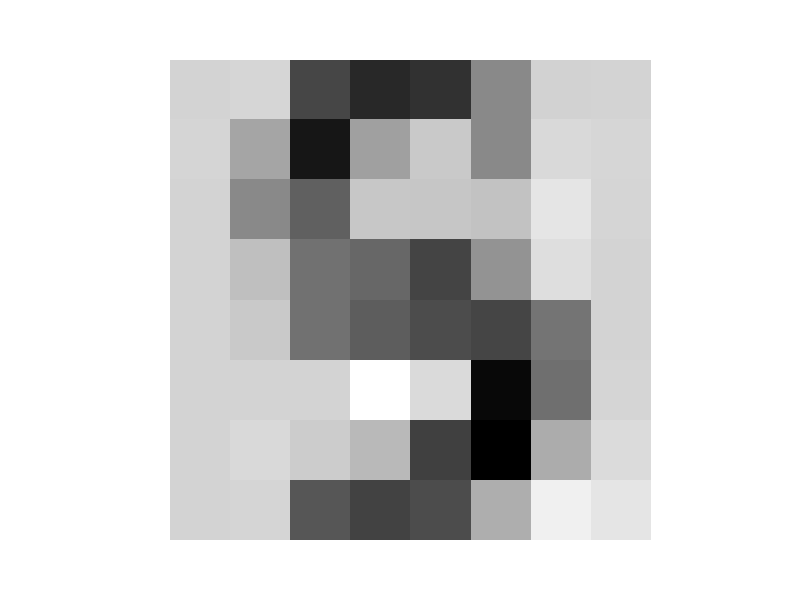}
\end{subfigure}%
\hfill
\begin{subfigure}[b]{\imsize}
	\centering
	\includegraphics[width=\imsize]{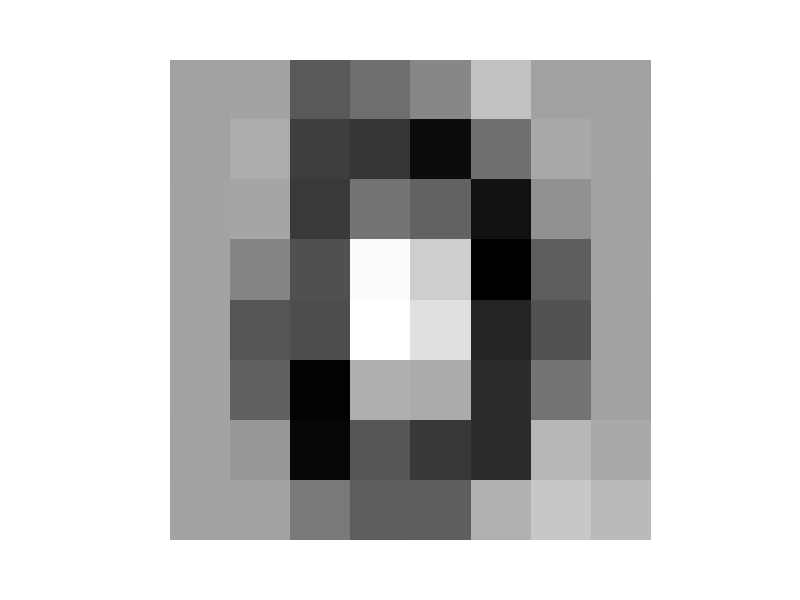}
\end{subfigure}%
\hfill
\begin{subfigure}[b]{\imsize}
	\centering
	\includegraphics[width=\imsize]{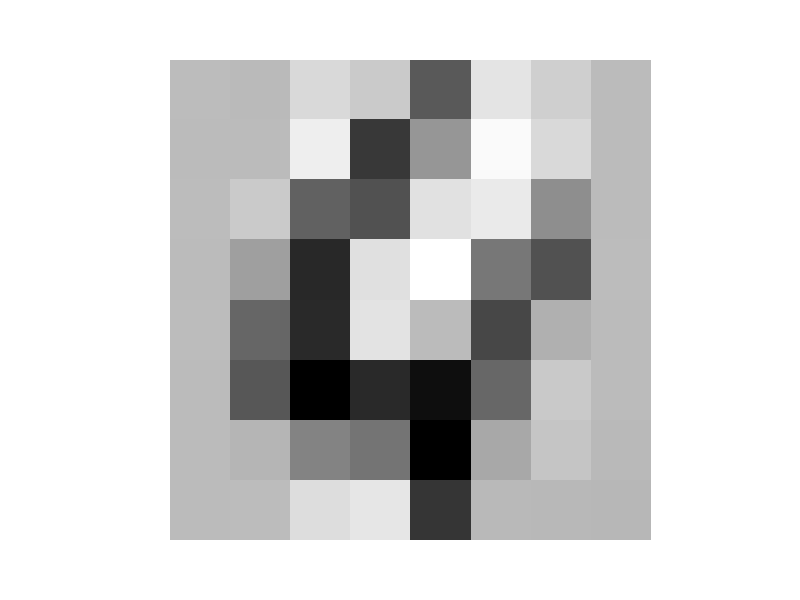}
\end{subfigure}%
\hfill
\begin{subfigure}[b]{\imsize}
	\centering
	\includegraphics[width=\imsize]{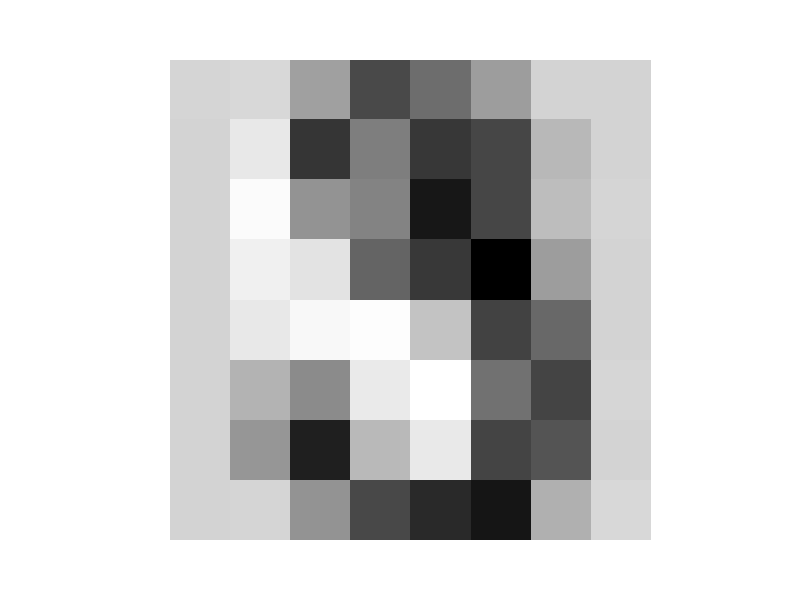}
\end{subfigure}%
\vspace{-4pt}
\caption{}
\label{fig:klr:20}
\end{subfigure}
~
\vline
~
\begin{subfigure}[b]{0.45\columnwidth}

\begin{subfigure}[b]{\imsize}
	\centering
	\includegraphics[width=\imsize]{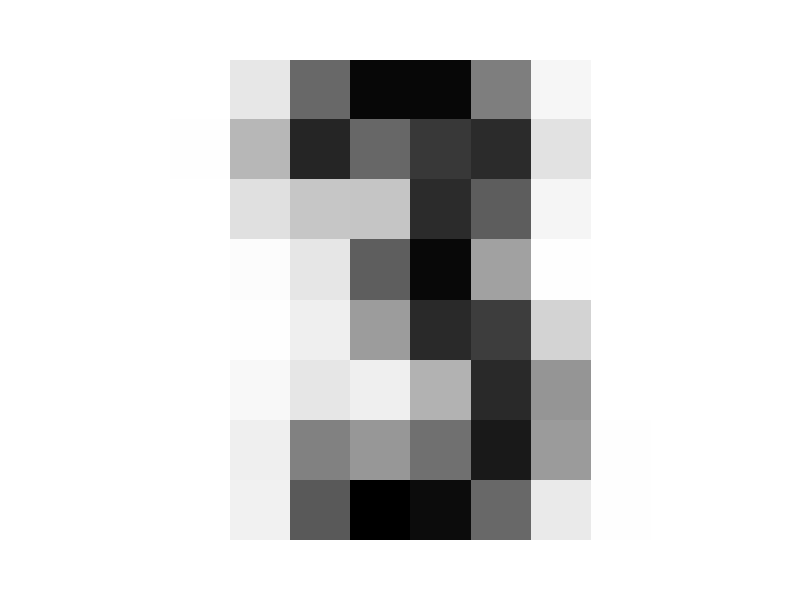}
\end{subfigure}%
\hfill
\begin{subfigure}[b]{\imsize}
	\centering
	\includegraphics[width=\imsize]{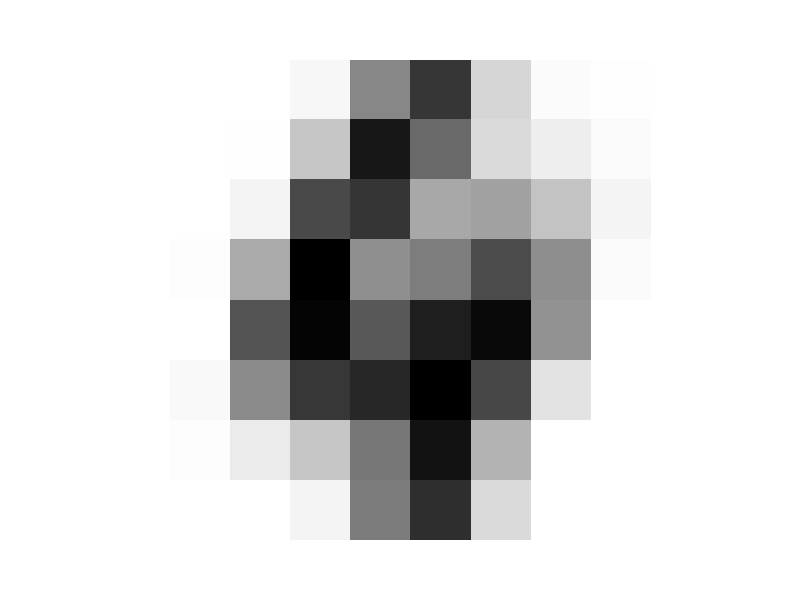}
\end{subfigure}%
\hfill
\begin{subfigure}[b]{\imsize}
	\centering
	\includegraphics[width=\imsize]{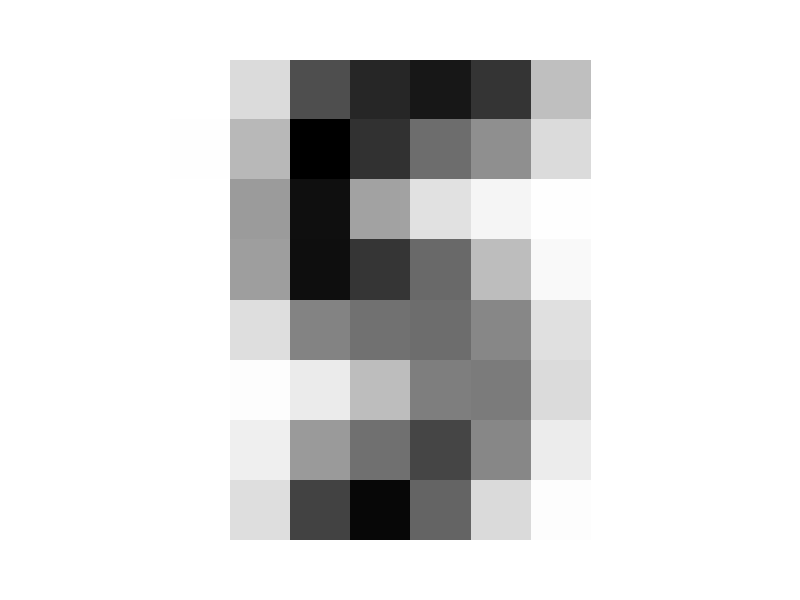}
\end{subfigure}%
\hfill
\begin{subfigure}[b]{\imsize}
	\centering
	\includegraphics[width=\imsize]{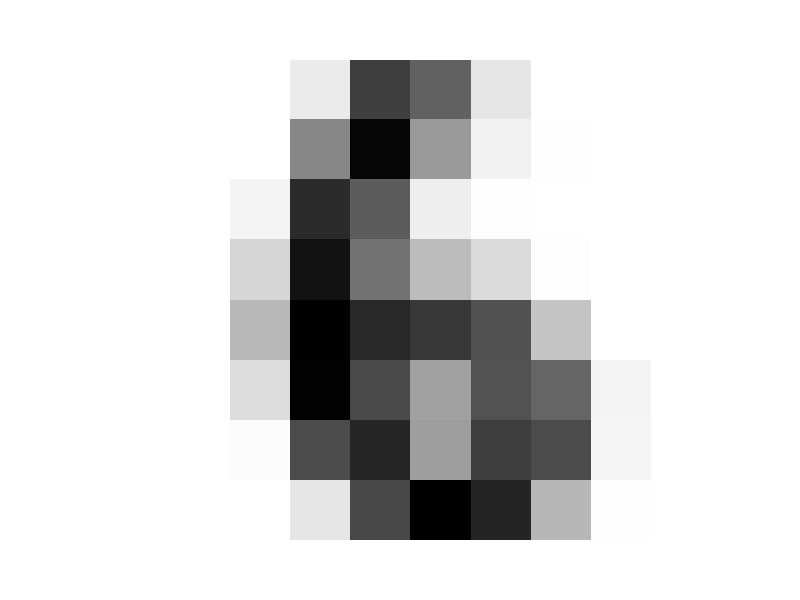}
\end{subfigure}%
\hfill
\begin{subfigure}[b]{\imsize}
	\centering
	\includegraphics[width=\imsize]{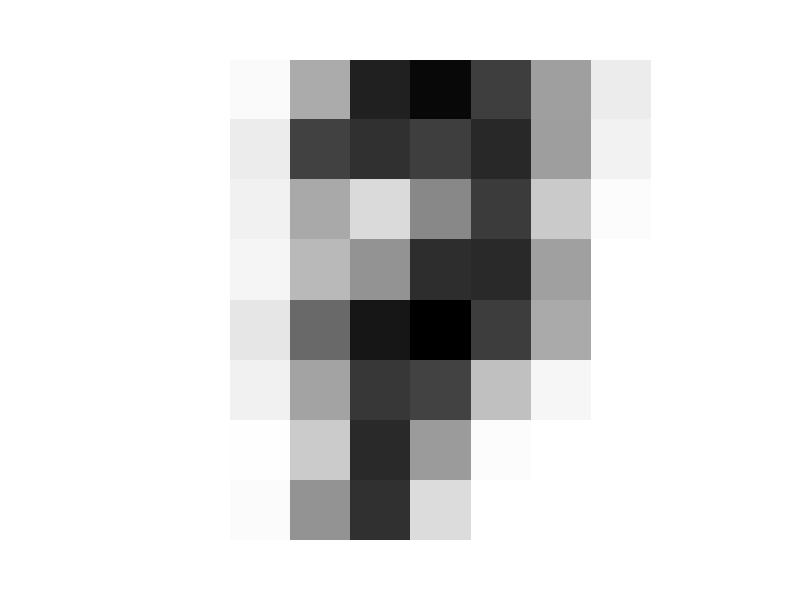}
\end{subfigure}%
\\
\begin{subfigure}[b]{\imsize}
	\centering
	\includegraphics[width=\imsize]{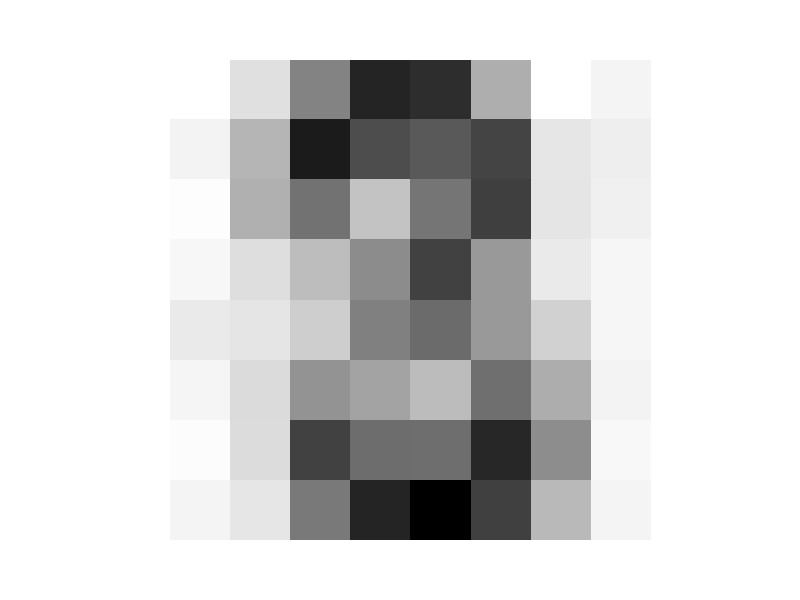}
\end{subfigure}%
\hfill
\begin{subfigure}[b]{\imsize}
	\centering
	\includegraphics[width=\imsize]{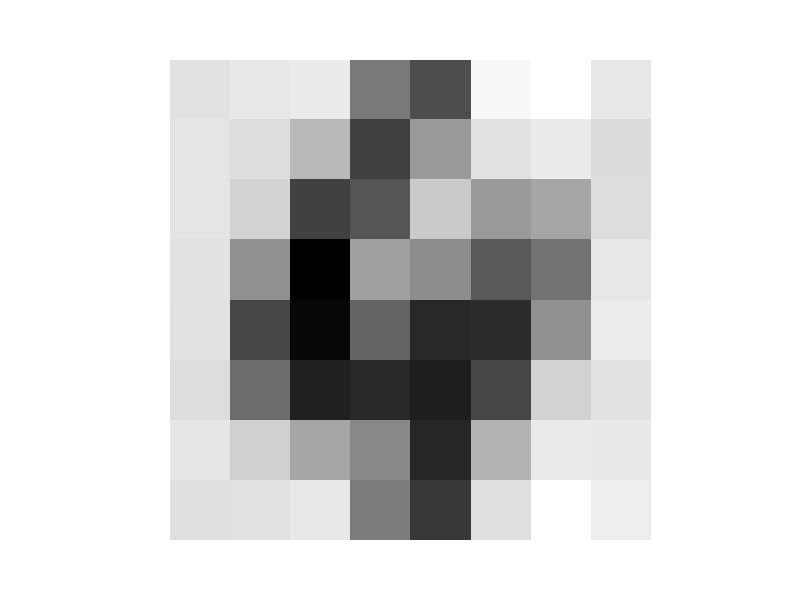}
\end{subfigure}%
\hfill
\begin{subfigure}[b]{\imsize}
	\centering
	\includegraphics[width=\imsize]{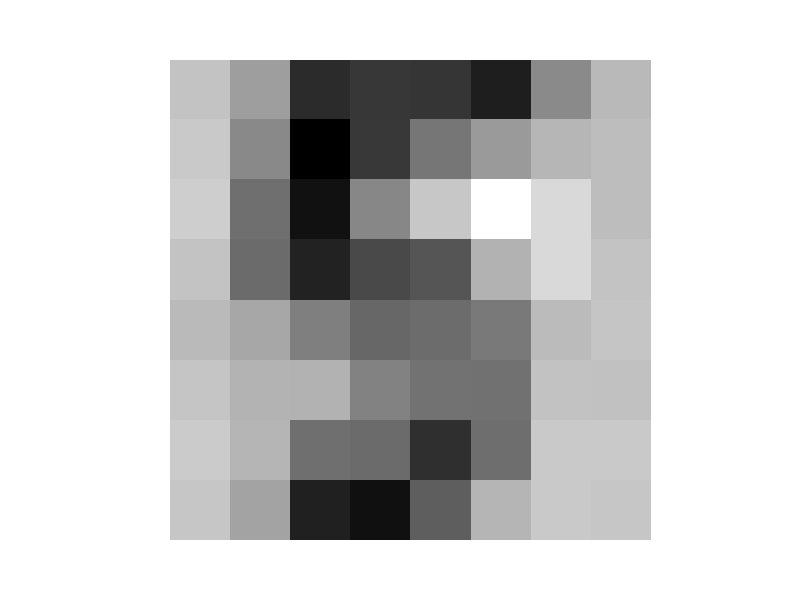}
\end{subfigure}%
\hfill
\begin{subfigure}[b]{\imsize}
	\centering
	\includegraphics[width=\imsize]{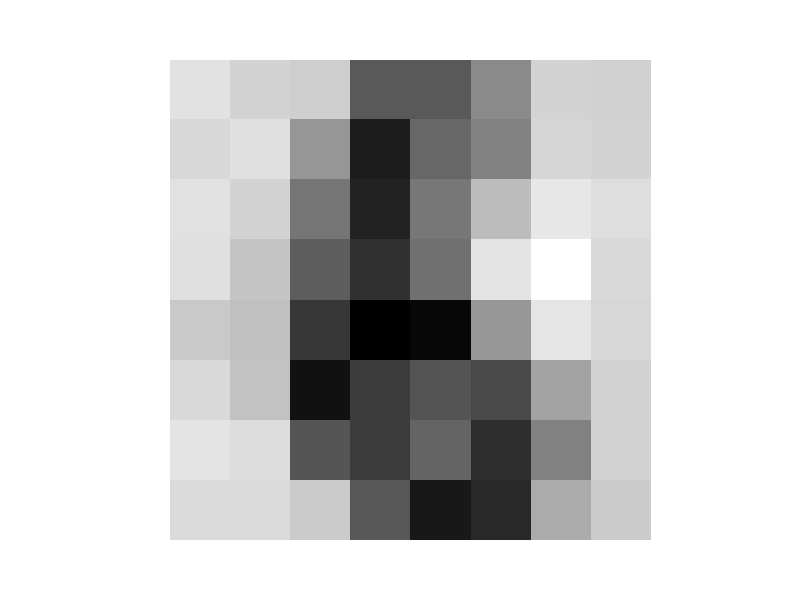}
\end{subfigure}%
\hfill
\begin{subfigure}[b]{\imsize}
	\centering
	\includegraphics[width=\imsize]{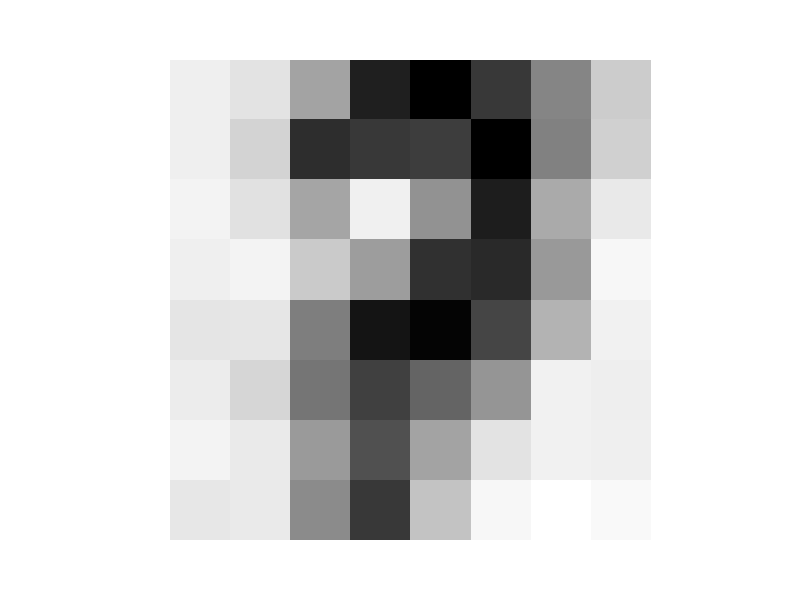}
\end{subfigure}%
\vspace{-4pt}
\caption{}
\label{fig:klr:all}
\end{subfigure}
\vspace{-9pt}
\caption{\footnotesize{\textbf{Training data leakage in KLR models.} 
(a) Displays $5$ of $20$ training samples used as representers in a KLR model (top) and $5$ of $20$ extracted representers (bottom).
(b) For a second model, shows the average of all $1{,}257$ representers that the model classifies as a $3,4,5,6$ or $7$ (top) and $5$ of $10$ extracted representers (bottom).}}
\label{fig:klr}
\end{figure}

A KLR model is a softmax model, where we replace the linear components $\wvec_i \cdot \xvec + \beta_i$ by a mapping $\sum_{r=1}^s \alpha_{i,r} K(\xvec, \xvec_r) + \beta_i$. Here, $K$ is a kernel function, and the \emph{representers} $\xvec_1, \dots, \xvec_s$ are a chosen subset of the training points~\cite{zhu01}. More details are in \apref{sec:model-details}.

Each sample $(\xvec, f(\xvec))$ from a KLR model yields $\numclasses$ equations over the parameters $\alphavec \in \R^{s \numclasses}, \betavec \in \R^\numclasses$ and the representers $\xvec_1, \dots, \xvec_s$. Thus, by querying the model, $\advA$ obtains a non-linear equation system, the solution of which leaks training data.
This assumes that $\advA$ knows the exact number $s$ of representers sampled from the data. However, we can relax this assumption: First, note that $f$'s outputs are unchanged by adding `extra' representers, with weights $\alpha=0$. Thus, over-estimating $s$ still results in a consistent system of equations, of which a solution is the model $f$, augmented with unused representers.
We will also show experimentally that training data may leak even if $\advA$ extracts a model $\fout$ with $s' \ll s$ representers.

We build two KLR models with a \emph{radial-basis function} (RBF) kernel for a data set of handwritten digits. We select $20$ random digits as representers for the first model, and all $1{,}257$ training points for the second. We extract the first model, assuming knowledge of $s$, by solving a system of $50{,}000$ equations in $1{,}490$ unknowns. We use the same approach as for MLPs, \ie logistic-loss minimization using gradient descent. We initialize the extracted representers to uniformly random vectors in $\fspace$, as we assume $\advA$ does not know the training data distribution. In \figref{fig:klr:20}, we plot $5$ of the model's representers from the training data, and the $5$ closest (in $l_1$ norm) extracted representers. The attack clearly leaks information on individual training points. We measure the attack's robustness to uncertainty about $s$, by attacking the second model with only $10$ local representers ($10{,}000$ equations in $750$ unknowns). \figref{fig:klr:all} shows the \emph{average} image of training points classified as a $3,4,5,6$ or $7$ by the target model $f$, along with $5$ extracted representers of $\fout$. Surprisingly maybe, the attack seems to be leaking the `average representor' of each class in the training data.

\input{inversion}

%% file: inversion.tex
\subsubsection{Model Inversion Attacks on Extracted Models}
\label{sec:inversion}

Access to a model may enable inference of privacy-damaging information,
particularly about the training
set~\cite{fredrikson14,fredrikson15,ateniese2015hacking}.  The \emph{model
inversion attack} explored by Fredrikson et al.~\cite{fredrikson15} uses access
to a classifier $f$ to find the input $\xvec_{\rm opt}$ that maximizes the class
probability for class $i$, \ie $\xvec_{\rm opt} = \argmax_{\xvec \in \fspace}
f_i(\xvec)$. This was shown to allow recovery of recognizable images of training
set members' faces when $f$ is a facial recognition model.

Their attacks work best in a \emph{white-box} setting, where the attacker knows $f$ and its
parameters.  Yet, the authors also note that in a
black-box setting, remote queries to a prediction API, combined with numerical
approximation techniques, enable successful, albeit much less
efficient, attacks. Furthermore, their black-box attacks inherently require $f$ to be
queried \emph{adaptively}. 
They leave as an open question making black-box attacks more
efficient.

We explore composing an attack that first 
attempts to \emph{extract} a model $\fout
\approx f$, and then uses it with the \cite{fredrikson15} white-box inversion attack.
Our extraction techniques replace adaptive queries with a non-adaptive
``batch'' query to $f$, followed by local computation. We show that 
extraction plus inversion can require fewer queries and less time than performing
 black-box inversion directly.

As a case study, we use the softmax
model from~\cite{fredrikson15}, trained over the AT\&T Faces
data~\cite{attfaces}. The data set consists of images of faces ($92\times112$ pixels) of $40$ people. The black-box attack
from~\cite{fredrikson15} needs about $20{,}600$ queries to reconstruct a
recognizable face for a single training set individual.  
Reconstructing the faces of all $40$ individuals
would require around $800{,}000$ online queries.

The trained softmax model is much larger than those considered in \secref{sec:eqsolve}, with $412{,}160$ unknowns ($\fdim=10{,}304$ and $\numclasses=40$). 
We solve an under-determined system with $41{,}216$ equations 
(using gradient descent with $200$ epochs), and recover a model $\fout$ achieving
$\errortesttv, \erroruniftv$ in the order of $10^{-3}$.  Note that the number of 
model parameters to extract is linear in the number of people $c$, whose faces
we hope to recover.
By using $\fout$ in
white-box model inversion attacks, we obtain results that are visually
indistinguishable from the ones obtained using the true $f$.  Given the
extracted model $\fout$, we can recover all $40$ faces using white-box
attacks, incurring around $20\times$ fewer remote queries to $f$ than with $40$
black-box attacks.
 
For black-box attacks, the authors of~\cite{fredrikson15} estimate a query
latency of $70$ milliseconds (a little less than in our own measurements of ML services, see Table~\ref{tab:result-summary}). Thus, it takes $24$ minutes to recover a single face (the inversion attack runs in seconds), and $16$ hours to recover all $40$ images. In contrast, solving the large equation system
underlying our model-extraction attack took $10$ hours. The $41{,}216$ online queries would take under one hour if executed sequentially and even less with a batch query. The cost of the $40$ local white-box attacks is negligible.

Thus, if the goal is to reconstruct faces for all $40$ training individuals,
performing model inversion over a previously extracted model results in an attack that is both faster and requires $20\times$ fewer online queries.

%% file: dectrees.tex
\subsection{Decision Tree Path-Finding Attacks}
\label{sec:dectrees}

Contrary to logistic models, decision trees do not compute class probabilities as a continuous function of their input. Rather, decision trees partition the input space into discrete regions, each of which is assigned a label and confidence score.
We propose a new \emph{path-finding} attack, that exploits API particularities to extract the `decisions' taken by a tree when classifying an input.

Prior work on decision tree extraction~\cite{bellare1992technique,kushilevitz1993learning,blum97} has focused on trees with Boolean features and outputs. While of theoretical importance, such trees have limited practical use.
Kushilevitz and Mansour~\cite{kushilevitz1993learning} showed that Boolean trees can be extracted using membership queries (arbitrary queries for class labels), but their algorithm does not extend to more general trees. Here, we propose attacks that exploit ML API specificities, and that apply to decision tree models used in MLaaS platforms.

Our tree model, defined formally in \apref{sec:model-details}, allows for binary and multi-ary splits over categorical features, and binary splits over numeric features.
Each leaf of the tree is labeled with a class label and a confidence score.
We note that our attacks also apply (often with better results) to \emph{regression trees}. In regression trees, each leaf is labeled with a real-valued output and confidence.

The key idea behind our attack is to use the rich information provided by APIs on a prediction query, as a \emph{pseudo-identifier} for the \emph{path} that the input traversed in the tree.
By varying the value of each input feature, we then find the predicates to be satisfied, for an input to follow a given path in the tree.
We will also exploit the ability to query \emph{incomplete inputs}, in which each feature $x_i$ is chosen from a space $\fspace_i \cup \set{\bot}$, where $\bot$ encodes the absence of a value. One way of handling such inputs (\cite{bigml, saar07}) is to label each node in the tree with an output value. On an input, we traverse the tree until we reach a leaf or an internal node with a split over a missing feature, and output that value of that leaf or node.

We formalize these notions by defining \emph{oracles} that $\advA$ can query to obtain an identifier for the leaf or internal node reached by an input. In practice, we instantiate these oracles using prediction API peculiarities.

\begin{definition}[Identity Oracles]
Let each node $v$ of a tree $T$ be assigned some identifier $\id_v$. A leaf-identity oracle $\oracle$ takes as input a query $\xvec \in \fspace$ and returns the identifier of the leaf of the tree $T$ that is reached on input $\xvec$.

A node-identity oracle $\oracle_{\bot}$ takes as input a query $\xvec \in \fspace_1 \cup \set{\bot} \times \cdots \times \fspace_d \cup \set{\bot}$ and returns the identifier of the node or leaf of $T$ at which the tree computation halts.
\end{definition}

\subsubsection{Extraction Algorithms}

\begin{algorithm}[t]
{\fontsize{8}{8.7}\selectfont
\begin{algorithmic}[1]
	\State $\xvec_{\text{init}} \gets \set{x_1, \dots, x_\fdim}$ \Comment{random initial query}
	\State $Q \gets \set{\xvec_\text{init}}$ \Comment{Set of unprocessed queries}
	\State $P \gets \set{}$ \Comment{Set of explored leaves with their predicates}
	\While{$Q$ not empty}
		\State $\xvec \gets Q.$\Call{pop}{\null}
		\State $\id \gets \oracle(\xvec)$ \Comment{Call to the leaf identity oracle}
		\If{$\id \in P$} \Comment{Check if leaf already visited} \label{alg:line:id-test}
			\Continue
		\EndIf
			\For{$1 \leq i \leq \fdim$} \Comment{Test all features}
				\If{\Call{is\_continuous}{$i$}}
					\For{$(\alpha, \beta] \in \Call{line\_search}{\xvec, i, \epsilon}$} \label{alg:line:ls}
						\If{$x_i \in (\alpha, \beta]$}
							\State $P[\id].\Call{add}{`x_i \in (\alpha, \beta]`}$ \Comment{Current interval}
						\Else
							\State $Q.$\Call{push}{$\xvec[i] \Rightarrow \beta$} \Comment{New leaf to visit}
						\EndIf
					\EndFor					
				\Else				
					\State $S, V \gets \Call{category\_split}{\xvec, i, \id}$ \label{alg:line:cs}
					\State $P[\id].\Call{add}{`x_i \in S`}$ \Comment{Values for current leaf}
					\For{$v \in V$}
						\State $Q.$\Call{push}{$\xvec[i] \Rightarrow v$} \Comment{New leaves to visit}
					\EndFor
				\EndIf
			\EndFor
	\EndWhile
\end{algorithmic}
}
\caption{\footnotesize{\textbf{The path-finding algorithm.} The notation $\id \gets \oracle(\xvec)$ means querying the leaf-identity oracle $\oracle$ with an input $\xvec$ and obtaining a response $\id$. By $\xvec[i] \Rightarrow v$ we denote the query $\xvec'$ obtained from $\xvec$ by replacing the value of $x_i$ by $v$.}}
\label{alg:extract}
\end{algorithm}

We now present our path-finding attack (\algref{alg:extract}), that assumes a leaf-identity oracle that returns \emph{unique} identifiers for each leaf. We will relax the uniqueness assumption further on.
The attack starts with a random input $\xvec$ and gets the leaf $\id$ from the
oracle. We then search for all constraints on $\xvec$
that have to be satisfied to remain in that leaf, using procedures \texttt{LINE\_SEARCH} (for continuous features) and
\texttt{CAT\_SPLIT} (for categorical features) described below.
From this information, we then create new queries for unvisited leaves. 
Once all leaves have been
found, the algorithm returns, for each leaf, the corresponding
constraints on $\xvec$. 
We analyze the algorithm's correctness
and complexity in \apref{sec:trees-analysis}.

We illustrate our algorithm with a toy example of a tree over continuous feature \emph{Size} and categorical feature \emph{Color} (see \figref{fig:treeContCat}). The current query is $\xvec = \{Size=50,\  Color=R\}$ and $\oracle(\xvec) = \id_2$. Our goal is two-fold: (1) Find the \emph{predicates} that $\xvec$ has to satisfy to end up in leaf $\id_2$ (\ie $Size \in (40, 60],\  Color=R$), and (2) create new inputs $\xvec'$ to explore other paths in the tree.

The \texttt{LINE\_SEARCH} procedure (line~\ref{alg:line:ls}) tests \emph{continuous features}.
We start from bounds on the range of a feature $\fspace_i = [a,b]$. In our example, we have $\text{\emph{Size}} \in [0, 100]$. We set the value of \emph{Size} in $\xvec$ to $0$ and $100$, query $\oracle$, and obtain $\id_1$ and $\id_5$. As the $\id$s do not match, a split on \emph{Size} occurs on the path to $\id_2$. With a binary search over feature \emph{Size} (and all other features in $\xvec$ fixed), we find all intervals that lead to different leaves, \ie $[0, 40]$, $(40, 60]$, $(60, 100]$. From these intervals, 
we find the predicate for the current leaf (\ie $Size \in (40, 60]$) and build queries to explore new tree paths.
To ensure termination of the line search, we specify some \emph{precision} $\epsilon$. If a split is on a threshold $t$, we find the value $\tilde{t}$ that is the unique multiple of $\epsilon$ in the range $(t-\epsilon, t]$. For values $x_i$ with granularity $\epsilon$, splitting on $\tilde{t}$ is then equivalent to splitting on $t$.

The \texttt{CATEGORY\_SPLIT} procedure (line~\ref{alg:line:cs}) finds splits on \emph{categorical features}.
In our example, we vary the value of \emph{Color} in $\xvec$ and query $\oracle$ to get a leaf $\id$ for each value. We then build a set $S$ of values that lead to the current leaf, \ie $S=\set{\texttt{R}}$, and a set $V$ of values to set in $\xvec$ to explore other leaves (one representative per leaf). In our example, we could have $V=\set{\texttt{B},\texttt{G},\texttt{Y}}$ or $V=\set{\texttt{B},\texttt{G},\texttt{O}}$.

Using these two procedures, we thus find the predicates defining the path to leaf $\id_2$, and generate new queries $\xvec'$ for unvisited leaves of the tree.

\begin{figure}[t]
\footnotesize
\centering
\begin{tikzpicture}
  [
  	level 1/.style={sibling distance=45mm, level distance=1em},
  	level 2/.style={sibling distance=30mm, level distance=2.2em},
  	level 3/.style={sibling distance=25mm, level distance=2.2em},
  	level 4/.style={sibling distance=15mm, level distance=3em},
    edge from parent/.style = {draw, -latex},
    every node/.style       = {font=\scriptsize}
  ]
  \node [inner] {Color}
	child[edge from parent/.style={OliveGreen,very thick,draw, -latex}] { node [inner] {Size}	  
		child[edge from parent/.style={black, thin, draw, -latex}] { node [leaf] {$\id_1$} 
			edge from parent node [left, yshift=4pt,xshift=0pt] {$\leq 40$}}
  		child[edge from parent/.style={OliveGreen,very thick,draw, -latex}] { node [inner] {Size}
  			child[edge from parent/.style={OliveGreen,very thick,draw, -latex}] { node [inner] {Color}
				child[edge from parent/.style={OliveGreen,very thick,draw, -latex}] { node [leafShade] {$\id_2$} 
					edge from parent node [left, yshift=3pt, xshift=2pt] {$=\texttt{R}$}
					}
  				child[edge from parent/.style={black, thin, draw, -latex}] { node [leaf] {$\id_3$} 
  					edge from parent node [right, yshift=1pt, xshift=-2pt] {$=\texttt{B}$}}
  				child[edge from parent/.style={black, thin, draw, -latex}] { node [leaf] {$\id_4$} 
  					edge from parent node [right, yshift=3pt, xshift=-2pt] {$=\texttt{G}$}}			
  				edge from parent node [left, yshift=5pt,xshift=4pt] {$\leq 60$}}	
  			child[edge from parent/.style={black, thin,draw, -latex}] { node [leaf] {$\id_5$} 
				edge from parent node [right, yshift=5pt,xshift=-4pt] {$> 60$}}	
  			edge from parent node [right, yshift=4pt,xshift=0pt] {$> 40$}}
  		edge from parent node [left, yshift=6pt, xshift=8pt] {$\in \set{\texttt{R}, \texttt{B}, \texttt{G}}$}}
  	child[edge from parent/.style={black, thin, draw, -latex}] { node [leaf] {$\id_6$} 
  		edge from parent node [right, yshift=6pt, xshift=-8pt] {$\in \set{\texttt{Y}, \texttt{O}}$}};
\end{tikzpicture}

\vspace{-8pt}
\caption{\footnotesize{\textbf{Decision tree over features Color and Size.} Shows the path (thick green) to leaf $\id_2$ on input $\xvec = \{Size=50,\  Color=R\}$.
}}
\label{fig:treeContCat}
\end{figure}

\paragraph{A top-down approach.}
We propose an empirically more efficient \emph{top-down} algorithm that exploits queries over partial inputs. It extracts the tree `layer by layer', starting at the root: We start with an empty query (all features set to $\bot$) and get the root's $\id$ by querying $\oracle_{\bot}$. We then set each feature in turn and query $\oracle$ again. For exactly one feature (the root's splitting feature), the input will reach a different node. With similar procedures as described previously, we extract the root's splitting criterion, and recursively search lower layers of the tree. 


\paragraph{Duplicate identities.}
As we verify empirically, our attacks are resilient to some nodes or leaves sharing the same $\id$. We can modify line~\ref{alg:line:id-test} in \algref{alg:extract} to detect $\id$ duplicates, by checking not only whether a leaf with the current $\id$ was already visited, but also whether the current query violates that leaf's predicates. The main issue with duplicate $\id$s comes from the \texttt{LINE\_SEARCH} and \texttt{CATEGORY\_SPLIT} procedures: if two queries $\xvec$ and $\xvec'$ differ in a single feature and reach different leaves with the same $\id$, the split on that feature will be missed.

\subsubsection{Attack Evaluation}

\begin{table}[t]
\def\arraystretch{0.85}
\center
\footnotesize
\begin{tabularx}{\columnwidth}{@{} l r  *{2}{>{\raggedleft\arraybackslash}X @{}}}
\textbf{Data set} & \textbf{\# records} & \textbf{\# classes} & \textbf{\# features} \\
\toprule
IRS Tax Patterns & $191{,}283$ & $51$ & $31$ \\
Steak Survey & $430$ & $5$ & $12$ \\
GSS Survey & $51{,}020$ & $3$ & $7$ \\
Email Importance & $4{,}709$ & $2$ & $14$\\
Email Spam &  $4{,}601$ & $2$ & $46$ \\
German Credit & $1{,}000$ & $2$ & $11$ \\
\midrule
Medical Cover & $163{,}065$ & $\outspace = \R$ & $13$\\
Bitcoin Price & $1{,}076$ & $\outspace = \R$ & $7$ \\
\bottomrule
\end{tabularx}
\vspace{-10pt}
\caption{\footnotesize{\textbf{Data sets used for decision tree extraction.} Trained trees for these data sets are available in BigML's public gallery. 
The last two data sets are used to train regression trees.}}
\label{tab:benchmark-trees}
\end{table}

\begin{table*}[t]
\footnotesize
\def\arraystretch{0.85}
\begin{tabularx}{\textwidth}{@{} l r r r @{\extracolsep{\fill}} r r r r r r @{}} 
	  &			&		 &				   & \multicolumn{3}{c}{\textbf{Without incomplete queries}} & \multicolumn{3}{c}{\textbf{With incomplete queries}}\vspace{-1pt}\\
\textbf{Model} & \textbf{Leaves} & \textbf{Unique IDs} & \textbf{Depth} & $1-\errortest$ & $1-\errorunif$ & {Queries} & $1-\errortest$ & $1-\errorunif$ & {Queries} \\[1pt]
\toprule
IRS Tax Patterns 	& $318$ & $318$ &  $8$ & $100.00\%$ & $100.00\%$ & $101{,}057$ & $100.00\%$ & $100.00\%$ & $29{,}609$ \\
Steak Survey 		& $193$ &  $28$ & $17$ & $92.45\%$ & $86.40\%$ & $3{,}652$ & $100.00\%$ & $100.00\%$ & $4{,}013$ \\
GSS Survey 			& $159$ & $113$ &  $8$ & $99.98\%$ & $99.61\%$ & $7{,}434$ & $100.00\%$ & $99.65\%$ & $2{,}752$ \\
Email Importance 	& $109$ &  $55$ & $17$ & $99.13\%$ & $99.90\%$ & $12{,}888$ & $99.81\%$ & $99.99\%$ & $4{,}081$ \\
Email Spam 			& $219$ &  $78$ & $29$ & $87.20\%$ & $100.00\%$ & $42{,}324$ & $99.70\%$ & $100.00\%$ & $21{,}808$ \\
German Credit 		&  $26$ &  $25$ & $11$ & $100.00\%$ & $100.00\%$ & $1{,}722$ & $100.00\%$ & $100.00\%$ & $1{,}150$ \\
\midrule
Medical Cover 		&  $49$ &  $49$ & $11$ & $100.00\%$ & $100.00\%$ & $5{,}966$ & $100.00\%$ & $100.00\%$ & $1{,}788$ \\
Bitcoin Price 		& $155$ & $155$ &  $9$ & $100.00\%$ & $100.00\%$ & $31{,}956$ & $100.00\%$ & $100.00\%$ & $7{,}390$ \\
\bottomrule
\end{tabularx}
\vspace{-10pt}
\caption{\footnotesize{\textbf{Performance of extraction attacks on public models from BigML.} For each model, we report the number of leaves in the tree, the number of unique identifiers for those leaves, and the maximal tree depth. The chosen granularity $\epsilon$ for continuous features is $10^{-3}$.}}
\label{tab:tree_results}
\end{table*}

Our tree model (see \apref{sec:model-details}) is the one used by BigML.  Other ML services use similar tree models. For our experiments, we downloaded eight public decision trees from BigML (see \tabref{tab:benchmark-trees}), and queried them locally using available API bindings. More details on these models are in \apref{sec:datasets}.
We show \emph{online} extraction attacks on black-box models from BigML in \secref{sec:real-world}.

To emulate black-box model access, we first issue online queries to BigML, to determine the information contained in the service's responses. We then simulate black-box access locally, by discarding any extra information returned by the local API. Specifically, we make use of the following fields in query responses:

\begin{newitemize}
\item \textbf{Prediction.} This entry contains the predicted class label (classification) or real-valued output (regression).
\item \textbf{Confidence.}  For classification and regression trees, BigML computes confidence scores based on a confidence interval for predictions at each node~\cite{bigml}. The prediction and confidence value constitute a node's $\id$.
\item \textbf{Fields.}  Responses to black-box queries contain a `fields' property, that lists all features that appear either in the input query or on the path traversed in the tree. If a partial query $\xvec$ reaches an internal node $v$, this entry tells us which feature $v$ splits on (the feature is in the `fields' entry, but not in the input $\xvec$). We make use of this property for the top-down attack variant.
\end{newitemize}

\tabref{tab:tree_results} displays the results of our attacks. For each tree, we give its number of leaves, the number of unique leaf $\id$s, and the tree depth. We display the success rate for \algref{alg:extract} and for the ``top-down'' variant with incomplete queries. Querying partial inputs vastly improves our attack: we require far less queries (except for the Steak Survey model, where \algref{alg:extract} only visits a fraction of all leaves and thus achieves low success) and achieve higher accuracy for trees with duplicate leaf $\id$s. 
As expected, both attacks achieve perfect extraction when all leaves have unique $\id$s. While this is not always the case for classification trees, it is far more likely for regression trees, where both the label and confidence score take real values. Surprisingly maybe, the top-down approach also fully extracts some trees with a large number of duplicate leaf $\id$s.
The attacks are also efficient: The top-down approach takes less than $10$ seconds to extract a tree, and \algref{alg:extract} takes less than $6$ minutes for the largest tree. For online attacks on ML services, discussed next, this cost is trumped by the delay for the inherently adaptive prediction queries that are issued.

%% file: ftransforms.tex
\section{Online Model Extraction Attacks}
\label{sec:real-world}

In this section, we showcase \emph{online} model extraction attacks against two ML services: BigML and Amazon. For BigML, we focus on extracting models set up by a user, who wishes to charge for predictions. For Amazon, our goal is to extract a model trained by ourselves, to which we only get black-box access. Our attacks only use exposed APIs, and do not in any way attempt to bypass the services' authentication or access-control mechanisms. We only attack models trained in our own accounts.

\subsection{Case Study 1: BigML}

BigML currently only allows monetization of decision trees~\cite{bigml}. We train a tree on the \emph{German Credit} data, and set it up as a black-box model. The tree has $26$ leaves, two of which share the same label and confidence score. From another account, we extract the model using the two attacks from~\secref{sec:dectrees}. 
We first find the tree's number of features, their type and their range, from BigML's public gallery.
Our attacks (\algref{alg:extract} and the top-down variant) extract an exact description of the tree's paths, using respectively $1{,}722$ and $1{,}150$ queries. Both attacks' duration ($1{,}030$ seconds and $631$ seconds) is dominated by query latency ($\approx500\text{ms}/\text{query}$). The monetary cost of the attack depends on the per-prediction-fee set by the model owner. In any case, 
a user who wishes to make more than $1{,}150$ predictions has economic incentives to run an extraction attack.

\subsection{Case Study 2: Amazon Web Services}

\begin{table}
\def\arraystretch{0.85}
\center\footnotesize
\begin{tabularx}{\columnwidth}{@{} l c c *{3}{>{\raggedleft\arraybackslash}X} @{}}
\textbf{Model} & \textbf{OHE} &\textbf{Binning}
& \textbf{Queries} & \textbf{Time (s)} & \textbf{Price ($\$$)}\\
\toprule
Circles & - & Yes & $278$ & $28$ & $0.03$\\
Digits & - & No & $650$ & $70$ & $0.07$\\
Iris & - & Yes & $644$ & $68$ & $0.07$\\
Adult & Yes & Yes & $1{,}485$ & $149$ & $0.15$\\
\bottomrule
\end{tabularx}
\vspace{-10pt}
\caption{\footnotesize{\textbf{Results of model extraction attacks on Amazon.} OHE stands for one-hot-encoding. The reported query count is the number used to find quantile bins (at a granularity of $10^{-3}$), plus those queries used for equation-solving. Amazon charges $\$0.0001$ per prediction~\cite{aws}.}}
\label{tab:results-aws}
\end{table}

Amazon uses logistic regression for classification, and provides black-box-only access to trained models~\cite{aws}. By default, Amazon uses two feature extraction techniques: (1) Categorical features are \emph{one-hot-encoded}, \ie the input space $\ispace_i = \Ints_k$ is mapped to $k$ binary features encoding the input value. (2) \emph{Quantile binning} is used for numeric features. The training data values are split into $k$-quantiles ($k$ equally-sized bins), and the input space $\ispace_i = [a,b]$ is mapped to $k$ binary features encoding the bin that a value falls into.
Note that $\abs{\fspace} > \abs{\ispace}$, \ie $\ext$ increases the number of features. If $\advA$ reverse-engineers $\ext$, she can query the service on samples $M$ in input space, compute $\xvec = \ext(M)$ locally, and extract $f$ in feature-space using equation-solving.

We apply this approach to models trained by Amazon. Our results are summarized in \tabref{tab:results-aws}. We first train a model with no categorical features, and quantile binning disabled (this is a manually tunable parameter), over the {Digits} data set. The attack is then identical to the one considered in \secref{sec:withconf:multiclass}: using $650$ queries to Amazon, we extract a model that achieves $\errortest = \errorunif = 0$.

We now consider models with feature extraction enabled.
We assume that $\advA$ knows the input space $\ispace$, but not the training data distribution. For one-hot-encoding, knowledge of $\ispace$ suffices to apply the same encoding locally. For quantile binning however, applying $\ext$ locally requires knowledge of the training data quantiles.
To reverse-engineer the binning transformation, we use line-searches similar to those we used for decision trees:
For each numeric feature, we search the feature's range in input space for thresholds (up to a granularity $\epsilon$) where $f$'s output changes. This indicates our value landed in an adjacent bin, with a different learned regression coefficient. 
Note that learning the bin boundaries may be interesting in its own right, as it leaks information about the training data distribution.
Having found the bin boundaries, we can apply both one-hot-encoding and binning locally, and extract $f$ over its feature space. As we are restricted to queries over $\ispace$, we cannot define an arbitrary system of equations over $\fspace$. Building a well-determined and consistent system can be difficult, as the encoding $\ext$ generates sparse inputs over $\fspace$. However, Amazon facilitates this process with the way it handles queries with \emph{missing features}: if a feature is omitted from a query, all corresponding features in $\fspace$ are set to $0$. For a linear model for instance, we can trivially re-construct the model by issuing queries with a single feature specified, such as to obtain equations with a single unknown in $\fspace$. 

We trained models for the {Circles}, {Iris} and {Adult} data sets, with Amazon's default feature-extraction settings. \tabref{tab:results-aws} shows the results of our attacks, for the reverse-engineering of $\ext$ and extraction of $f$. 
For binary models ({Circles} and {Adult}), we use $\fdim + 1$ queries to solve a
linear equation-system over $\fspace$. For models with $\numclasses > 2$ classes, we use $\numclasses \cdot
(\fdim + 1)$ queries.
In all cases, the extracted model matches $f$ on 100\% of tested inputs.
To optimize the query complexity, the queries we use to find quantile bins are re-used for equation-solving. As line searches require adaptive queries, we do not use batch predictions. However, even for the {Digits} model, we resorted to using real-time predictions, because of the service's significant overhead in evaluating batches.
For attacks that require a large number of non-adaptive queries, we expect batch predictions to be faster than real-time predictions.

\subsection{Discussion}

\paragraph{Additional feature extractors.}

In some ML services we considered, users may enable further feature extractors. A common transformation is feature scaling or normalization. If $\advA$ has access to training data statistics (as provided by BigML for instance), applying the transformation locally is trivial. More generally, for models with a linear input layer (\ie logistic regressions, linear SVMs, MLPs) the scaling or normalization can be seen as being applied to the learned weights, rather than the input features. We can thus view the composition $f \circ \ext$ as a model $f'$ that operates over the `un-scaled' input space $\ispace$ and extract $f'$ directly using equation-solving.

Further extractors include text analysis (\eg bag-of-words or n-gram models) and Cartesian products (grouping many features into one). We have not analyzed these in this work, but we believe that they could also be easily reverse-engineered, especially given some training data statistics and the ability to make incomplete queries.

\paragraph{Learning unknown model classes or hyper-parameters.}

For our online attacks, we obtained information about the model class of $f$, the enabled feature extraction $\ext$, and other hyper-parameters, directly from the ML service or its documentation. More generally, if $\advA$ does not have full certainty about certain model characteristics, it may be able to narrow down a guess to a small range. Model hyper-parameters for instance (such as the free parameter of an RBF kernel) are typically chosen through cross-validation over a default range of values.

Given a set of attack strategies with varying assumptions, $\advA$ can use a generic \emph{extract-and-test} approach: each attack is applied in turn, and evaluated by computing $\errortest$ or $\errorunif$ over a chosen set of points. The adversary succeeds if any of the strategies achieves a low error. Note that $\advA$ needs to interact with the model $f$ only once, to obtain responses for a chosen set of extraction samples and test samples, that can be re-used for each strategy.

Our attacks on Amazon's service followed this approach: We first formulated guesses for model characteristics left unspecified by the documentation (\eg we found no mention of one-hot-encoding, or of how missing inputs are handled). We then evaluated our assumptions with successive extraction attempts. Our results indicate that Amazon uses softmax regression and does not create binary predictors for missing values. Interestingly, BigML takes the 'opposite' approach (\ie BigML uses OvR regression and adds predictors for missing values).

%% file: labelonly.tex
\section{Extraction Given Class Labels Only}
\label{sec:labelonly}

The successful attacks given in Sections~\ref{sec:withconf} and \ref{sec:real-world} show the
danger of revealing confidence values. 
While current ML services have been designed to reveal rich information,
our attacks may suggest that returning only labels would be safer.
Here we explore model extraction in a setting with no confidence scores. We will discuss further countermeasures in~\secref{sec:countermeasures}.
We primarily focus on settings where $\advA$ can make \emph{direct} queries to an API, \ie queries for arbitrary inputs $\xvec \in \fspace$. We briefly discuss \emph{indirect} queries in the context of linear classifiers.

\paragraph{The Lowd-Meek attack.} 
We start with the prior work of Lowd and
Meek~\cite{lowd2005adversarial}. They present an attack on any linear classifier, 
assuming black-box oracle access with
membership queries that return just the predicted class label. 
A linear classifier is defined by a vector $\wvec \in
\R^\fdim$ and a constant $\beta \in \R$, and classifies an instance $\xvec$ as
\emph{positive} if $\wvec \cdot \xvec + \beta > 0$ and \emph{negative}
otherwise. SVMs with linear kernels and binary LRs are examples
of linear classifiers. Their attack uses line searches to find
points arbitrarily close to $f$'s decision boundary (points for which
$\wvec \cdot \xvec + \beta \approx 0$), and extracts $\wvec$ and $\beta$ from these samples.

This attack only works for linear binary models. We describe a straightforward 
extension to some non-linear models, such as 
polynomial kernel SVMs. Extracting a polynomial kernel SVM can be reduced to
extracting a linear SVM in the transformed feature space. Indeed, for any 
 kernel $K_{\text{poly}}(\xvec, \xvec')$$=$$(\xvec^T\cdot \xvec' + 1)^d$,
we can derive a projection function $\phi(\cdot)$, 
so that $K_{\text{poly}}(\xvec, \xvec')$$=$$\phi(\xvec)^T \cdot \phi(\xvec')$.
This transforms the kernel SVM into a linear one, since the
decision boundary now becomes $\wvec^{F} \cdot \phi(\xvec)
+ \beta = 0$ where $\wvec^F= \sum_{i=1}^t \alpha_i \phi(\xvec_i)$. We can 
use the Lowd-Meek attack to extract $\wvec^F$ and
$\beta$ as long as $\phi(\xvec)$ and its inverse are feasible to compute; this is unfortunately not 
the case for the more common RBF kernels.\footnote{We did explore using approximations of
$\phi$, but found that the adaptive re-training techniques discussed in this
section perform better.}

\paragraph{The retraining approach.} In addition to evaluating the
Lowd-Meek attack against ML APIs, we introduce a number of other approaches
based on the broad strategy of re-training a model locally, given input-output
examples. 
Informally, our hope is that by extracting a model that achieves
low \emph{training error} over the queried samples, we would effectively
approximate the target model's decision boundaries.
We consider three retraining strategies, described below.
We apply these to the model classes that we previously 
extracted using equation-solving attacks, as well as to SVMs.\footnote{We do not expect retraining attacks to work well for decision trees, because of the greedy approach taken by learning algorithms. We have not evaluated extraction of trees, given class labels only, in this work.}


\begin{newenum}
\item \textbf{Retraining with uniform queries.} This baseline strategy simply consists in sampling $m$ points $\xvec_i \in \fspace$ uniformly at random, querying the oracle, and training a model $\fout$ on these samples.
\item \textbf{Line-search retraining.} This strategy can be seen as a model-agnostic generalization of the Lowd-Meek attack. It issues $m$ adaptive queries to the oracle using line search techniques, to find samples close to the decision boundaries of $f$. A model $\fout$ is then trained on the $m$ queried samples.
\item \textbf{Adaptive retraining.} This strategy applies techniques from active learning~\cite{active, cohn1994improving}. For some number $r$ of rounds and a query budget $m$, it first queries the oracle on $\frac{m}{r}$ uniform points, and trains a model $\fout$. 
Over a total of $r$ rounds, it then selects $\frac{m}{r}$ new points, along the decision boundary of $\fout$ (intuitively, these are points $\fout$ is \emph{least certain} about), and sends those to the oracle before retraining $\fout$. 
\end{newenum}


\subsection{Linear Binary Models}
\label{sec:labelonly:linear}

\begin{figure}[t]
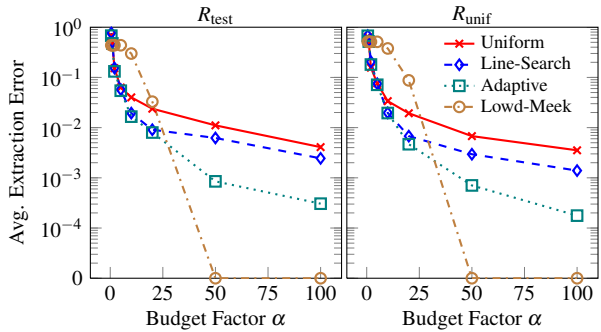

   \footnotesize
   \centering
   \hspace{-5pt}
   \include{data/regression/bin/full_plot}
   \vspace{-10pt}
   \caption{\footnotesize{\textbf{Average error of extracted linear models.} Results are for different extraction strategies applied to models trained on all binary data sets from \tabref{tab:benchmark}. The left shows $\errortest$ and the right shows $\errorunif$.}}
   \label{fig:results:linear}%
\end{figure}

We first explore how well the various approaches work in settings where the Lowd-Meek attack can be applied. 
We evaluate their attack and our three retraining strategies for logistic regression models trained over the binary data sets shown in \tabref{tab:benchmark}.
These models have $d+1$ parameters, and we vary the query budget as $\alpha \cdot (d+1)$, for $0.5 \leq \alpha \leq 100$.
\figref{fig:results:linear} displays the average errors $\errortest$ and $\errorunif$ over all models, as a function of $\alpha$.

The retraining strategies that search for points near the decision boundary clearly perform better than simple uniform retraining. The adaptive strategy is the most efficient of our three strategies. For relatively low budgets, it even outperforms the Lowd-Meek attack. However, for budgets large enough to run line searches in each dimension, the Lowd-Meek attack is clearly the most efficient. 

For the models we trained, about $2{,}050$ queries on average, and $5{,}650$ at most, are needed to run the Lowd-Meek attack effectively. This is $50\times$ more queries than what we needed for equation-solving attacks. With $827$ queries on average, adaptive retraining yields a model $\fout$ that matches $f$ on over $99\%$ of tested inputs. Thus, even if an ML API only provides class labels, efficient extraction attacks on linear models remain possible.

We further consider a setting where feature-extraction (specifically one-hot-encoding of categorical features) is applied by the ML service, rather than by the user. $\advA$ is then limited to indirect queries in input space. Lowd and Meek~\cite{lowd2005adversarial} note that their extraction attack does not work in this setting, as $\advA$ can not run line searches directly over $\fspace$. In contrast, for the linear models we trained, we observed no major difference in extraction accuracy for the adaptive-retraining strategy, when limited to queries over $\ispace$. We leave an in-depth study of model extraction with indirect queries, and class labels only, for future work.

\subsection{Multiclass LR Models}
\label{sec:labelonly:multiclass}

The Lowd-Meek attack is not applicable in multiclass ($\numclasses > 2$) settings, even when the decision boundary is a combination of linear boundaries (as in multiclass regression)~\cite{nelson2012query,stevens2013hardness}.
We thus focus on evaluating the three retraining attacks we introduced, for the type of ML models we expect to find in real-world applications.

We focus on softmax models here, as softmax and one-vs-rest models have identical output behaviors when only class labels are provided: in both cases, the class label for an input $\xvec$ is given by $\argmax_i (\wvec_i \cdot \xvec + \beta_i)$. From an extractor's perspective, it is thus irrelevant whether the target was trained using a softmax or OvR approach.

\begin{figure}[t]
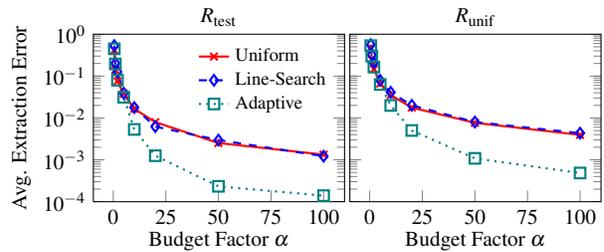

   \footnotesize
   \centering
   \hspace{-10pt}
   \include{data/regression/multiclass/full_plot}
   \vspace{-10pt}
   \caption{\footnotesize{\textbf{Average error of extracted softmax models.} Results are for three retraining strategies applied to models trained on all multiclass data sets from \tabref{tab:benchmark}. The left shows $\errortest$ and the right shows $\errorunif$.}}
   \label{fig:results:multiclass}%
\end{figure}

We evaluate our attacks on softmax models trained on the multiclass data sets shown in \tabref{tab:benchmark}. We again vary the query
budget as a factor $\alpha$ of the number of model parameters,
namely $\alpha\cdot \numclasses\cdot(d+1)$. Results are displayed in \figref{fig:results:multiclass}. We observe that the adaptive strategy clearly performs best and that the line-search strategy does not improve over uniform retraining, possibly because the line-searches have to be split across multiple decision-boundaries. We further note that all strategies achieve lower $\errortest$ than $\errorunif$. It thus appears that for the models we trained, points from the test set are on average `far' from the decision boundaries of $f$ (\ie the trained models separate the different classes with large margins). 

For all models, $100 \cdot \numclasses \cdot (d+1)$ queries resulted in extraction accuracy above $99.9\%$. This represents $26{,}000$ queries on average, and $65{,}000$ at the most (Digits data set). Our equation-solving attacks achieved similar or better results with 100$\times$ less queries.
Yet, for scenarios with high monetary incentives (\eg intrusion detector evasion), extraction attacks on MLR models may be attractive, even if APIs only provide class labels.

\begin{figure}[t]
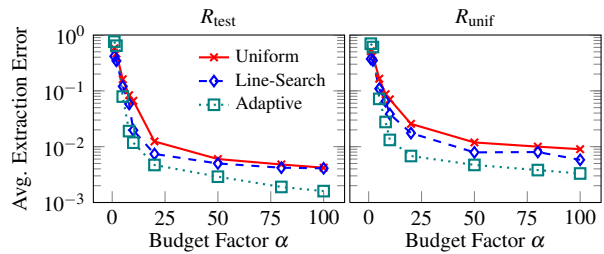

   \footnotesize
   \centering
   \hspace{-10pt}
   \include{data/svm/full_plot}
   \vspace{-12pt}
   \caption{\footnotesize{\textbf{Average error of extracted RBF kernel SVMs} Results are for three retraining strategies applied to models trained on all binary data sets from \tabref{tab:benchmark}. The left shows $\errortest$ and the right shows $\errorunif$.}}
   \label{fig:results:svm}%
\end{figure}

\subsection{Neural Networks}
\label{sec:labelonly:nnets}

We now turn to attacks on more complex deep neural networks. We expect these to be harder to retrain than multiclass regressions, as deep networks have more parameters and non-linear decision-boundaries. Therefore, we may need to find a large number of points close to a decision boundary in order to extract it accurately.

We evaluated our attacks on the multiclass models from \tabref{tab:benchmark}. For the tested query budgets, line-search and adaptive retraining gave little benefit over uniform retraining. For a budget of $100\cdot k$, where $k$ is the number of model parameters, we get $\errortest=99.16\%$ and $\errorunif=98.24\%$, using $108{,}200$ queries per model on average.
Our attacks might improve for higher budgets but it is unclear whether they would then provide any monetary advantage over using ML APIs in an honest way.

\subsection{RBF Kernel SVMs}
\label{sec:labelonly:svm}

Another class of nonlinear models that we consider are support-vector machines (SVMs) with radial-basis function (RBF) kernels. 
A kernel SVM first maps inputs into a higher-dimensional space, and then finds the hyperplane that maximally separates the two classes. As mentioned in \secref{sec:labelonly}, SVMs with polynomial kernels can be extracted using the Lowd-Meek attack in the transformed feature space. For RBF kernels, this is not possible because the transformed space has infinite dimension.

SVMs do not provide class probability estimates. Our only applicable attack is thus retraining. As for linear models, we vary the query budget as $\alpha \cdot (\fdim + 1)$, where $\fdim$ is the input dimension. We further use the \emph{extract-and-test} approach from \secref{sec:real-world} to find the value of the RBF kernel's \emph{hyper-parameter}. Results of our attacks are in \figref{fig:results:svm}. Again, we see that adaptive retraining performs best, even though the decision boundary to extract is non-linear (in input space) here.
Kernel SVMs models are overall harder to retrain than models with linear decision boundaries. Yet, for our largest budgets ($2{,}050$ queries on average), we do extract models with over 99\% accuracy, which may suffice in certain adversarial settings.

%% file: data/regression/bin/full_plot.tex
\pgfplotstableread[col sep=comma]{data/regression/bin/base_passive.dat}\databp
\pgfplotstableread[col sep=comma]{data/regression/bin/base_adapt-local.dat}\databal
\pgfplotstableread[col sep=comma]{data/regression/bin/base_adapt-oracle.dat}\databao
\pgfplotstableread[col sep=comma]{data/regression/bin/base_lowd-meek.dat}\datalm

\begin{tikzpicture}
  \begin{groupplot}[
  		group style={group size=2 by 1, 
  					horizontal sep = 2pt,
  	  				},
  	  	log origin=infty,
  	  	ymin=1e-5,
  	  	ymax=1,
  	  	width=0.62\columnwidth,
  	  	height=0.62\columnwidth,
  	  	xtick={0, 25, 50, 75, 100},
  	  	ytick={1e-5, 1e-4, 1e-3, 1e-2, 1e-1, 1},
  		cycle list name=list,
  		title style={at={(axis description cs:0.5,0.92)},anchor=south}, 
  		]
	
  \nextgroupplot[
  	  ymode=log, 
  	  xlabel=Budget Factor $\alpha$, 
  	  xlabel style={at={(axis description cs:0.5,-0.075)},anchor=south}, 
  	  xtick pos=left,
  	  ylabel=Avg. Extraction Error,
  	  title=$\errortest$,
  	  ylabel style={at={(axis description cs:0.15,.5)},anchor=south}, 
  	  yticklabels={$0$, $10^{-4}$, $10^{-3}$, $10^{-2}$, $10^{-1}$, $10^0$},
    ]
	
    \addplot
      table[x=Q_by_U,y expr=\thisrow{L_test_bar}+1e-5,y error=L_test_std, col sep=comma] {\databp};
    	
	    \addplot
      table[x=Q_by_U,y expr=\thisrow{L_test_bar}+1e-5,y error=L_test_std, col sep=comma] {\databao};
    	
    \addplot
      table[x=Q_by_U,y expr=\thisrow{L_test_bar}+1e-5,y error=L_test_std, col sep=comma] {\databal};
    
    \addplot
      table[x=Q_by_U,y expr=\thisrow{L_test_bar}+1e-5,y error=L_test_std, col sep=comma] {\datalm};

  \nextgroupplot[
  	ymode=log, 
  	xtick pos=left,
  	xlabel=Budget Factor $\alpha$, 
  	xlabel style={at={(axis description cs:0.5,-0.075)},anchor=south}, 
  	title=$\errorunif$, 
  	yticklabels={},
  	legend style = {font=\scriptsize, draw = none, fill=none, 
  					row sep=-2pt,
  					cells={anchor=west},
  	  				at={(0.35,1.01)}, anchor=north west,
  	  				 legend image code/.code={
  	      			 	\draw[mark repeat=2,mark phase=2]
						plot coordinates {
							(0cm,0cm)
							(0.2cm,0cm)        
							(0.4cm,0cm)         
						};}
					},
  ]
	
    \addplot
      table[x=Q_by_U,y expr=\thisrow{L_unif_bar}+1e-5,y error=L_unif_std, col sep=comma] {\databp};
    	\addlegendentry{Uniform};
    
    \addplot
      table[x=Q_by_U,y expr=\thisrow{L_unif_bar}+1e-5,y error=L_unif_std, col sep=comma] {\databao};
    \addlegendentry{Line-Search};
    
        \addplot
      table[x=Q_by_U,y expr=\thisrow{L_unif_bar}+1e-5,y error=L_unif_std, col sep=comma] {\databal};
    \addlegendentry{Adaptive};
    
    \addplot
      table[x=Q_by_U,y expr=\thisrow{L_unif_bar}+1e-5,y error=L_unif_std, col sep=comma] {\datalm};
    \addlegendentry{Lowd-Meek};
  
  \end{groupplot}    
\end{tikzpicture} 

%% file: data/regression/multiclass/full_plot.tex
\pgfplotstableread[col sep=comma]{data/regression/multiclass/softmax/base_passive.dat}\datasoftbp
\pgfplotstableread[col sep=comma]{data/regression/multiclass/softmax/base_adapt-local.dat}\datasoftbal
\pgfplotstableread[col sep=comma]{data/regression/multiclass/softmax/base_adapt-oracle.dat}\datasoftbao

\pgfplotstableread[col sep=comma]{data/regression/multiclass/ovr/base_passive.dat}\dataovrbp
\pgfplotstableread[col sep=comma]{data/regression/multiclass/ovr/base_adapt-local.dat}\dataovrbal
\pgfplotstableread[col sep=comma]{data/regression/multiclass/ovr/base_adapt-oracle.dat}\dataovrbao

\begin{tikzpicture}
  \begin{groupplot}[
  		group style={group size=2 by 1, 
  					horizontal sep = 2pt,
  					ylabels at=edge left,
  	  				},
  	  	log origin=infty,
  	  	ymin=1e-4,
  	  	ymax=1,
  	  	width=0.62\columnwidth,
  	  	height=0.48\columnwidth,
  	  	xtick={0, 25, 50, 75, 100},
  	  	ytick={1e-4, 1e-3, 1e-2, 1e-1, 1},
  		cycle list name=list,
  		title style={at={(axis description cs:0.5,0.91)},anchor=south}, 
  		]

  \nextgroupplot[
  	  ymode=log, 
  	  xlabel=Budget Factor $\alpha$, 
  	  xlabel style={at={(axis description cs:0.5,-0.1)},anchor=south}, 
  	  xtick pos=left,
  	  ylabel=Avg. Extraction Error,
  	  title=$\errortest$,
  	  ylabel style={at={(axis description cs:0.15,.5)},anchor=south}, 
	  legend style = {font=\scriptsize, draw = none, fill=none,
  					row sep=-1pt,
  	  				 cells={anchor=west},
  	      			 legend image code/.code={
  	      			 	\draw[mark repeat=2,mark phase=2]
						plot coordinates {
							(0cm,0cm)
							(0.2cm,0cm)        
							(0.4cm,0cm)         
						};}
					},
	]
	
    \addplot
      table[x=Q_by_U,y expr=\thisrow{L_test_bar}+1e-5,y error=L_test_std, col sep=comma] {\datasoftbp};
    \addlegendentry{Uniform};
    
    \addplot
      table[x=Q_by_U,y expr=\thisrow{L_test_bar}+1e-5,y error=L_test_std, col sep=comma] {\datasoftbao};
    \addlegendentry{Line-Search}; 
    
	\addplot
      table[x=Q_by_U,y expr=\thisrow{L_test_bar}+1e-5,y error=L_test_std, col sep=comma] {\datasoftbal};
    \addlegendentry{Adaptive};    
    
	\nextgroupplot[
  	ymode=log, 
  	xtick pos=left,
  	xlabel=Budget Factor $\alpha$, 
  	xlabel style={at={(axis description cs:0.5,-0.1)},anchor=south}, 
  	title=$\errorunif$, 
  	yticklabels={},
  ]
      
    \addplot
      table[x=Q_by_U,y expr=\thisrow{L_unif_bar}+1e-5,y error=L_unif_std, col sep=comma] {\datasoftbp};
    	
    \addplot
      table[x=Q_by_U,y expr=\thisrow{L_unif_bar}+1e-5,y error=L_unif_std, col sep=comma] {\datasoftbao};	
	
	\addplot
      table[x=Q_by_U,y expr=\thisrow{L_unif_bar}+1e-5,y error=L_unif_std, col sep=comma] {\datasoftbal};	
	
  \end{groupplot}
\end{tikzpicture} 

%% file: data/svm/full_plot.tex
\pgfplotstableread{data/svm/uniform.dat}\datauniform
\pgfplotstableread{data/svm/online.dat}\dataonline
\pgfplotstableread{data/svm/active.dat}\dataactive

\begin{tikzpicture}
  \begin{groupplot}[
  		group style={group size=2 by 1, 
  					horizontal sep = 2pt,
  					ylabels at=edge left,
  	  				},
  	  	log origin=infty,
  	  	ymin=1e-3,
  	  	ymax=1,
  	  	width=0.62\columnwidth,
  	  	height=0.48\columnwidth,
  	  	xtick={0, 25, 50, 75, 100},
  	  	ytick={1e-3, 1e-2, 1e-1, 1},
  		cycle list name=list,
  		title style={at={(axis description cs:0.5,0.91)},anchor=south}, 
  		]

  \nextgroupplot[
  	  ymode=log, 
  	  xlabel=Budget Factor $\alpha$, 
  	  xlabel style={at={(axis description cs:0.5,-0.1)},anchor=south}, 
  	  xtick pos=left,
  	  ylabel=Avg. Extraction Error,
  	  title=$\errortest$,
  	  ylabel style={at={(axis description cs:0.15,.5)},anchor=south}, 
  	  legend style = {font=\scriptsize, draw = none, fill=none,
  					 row sep=-1pt,
  	  				 cells={anchor=west},
  	      			 legend image code/.code={
  	      			 	\draw[mark repeat=2,mark phase=2]
						plot coordinates {
							(0cm,0cm)
							(0.2cm,0cm)        
							(0.4cm,0cm)         
						};}
					},
	]
	
    \addplot
      table[x=Q_by_U,y expr=\thisrow{L_test_bar},y error=L_test_std] {\datauniform};
    \addlegendentry{Uniform};
   
    \addplot
      table[x=Q_by_U,y expr=\thisrow{L_test_bar},y error=L_test_std] {\dataonline};
    \addlegendentry{Line-Search};
    
    \addplot
      table[x=Q_by_U,y expr=\thisrow{L_test_bar},y error=L_test_std] {\dataactive};
    \addlegendentry{Adaptive};

	\nextgroupplot[
  	ymode=log, 
  	xtick pos=left,
  	xlabel=Budget Factor $\alpha$, 
  	xlabel style={at={(axis description cs:0.5,-0.1)},anchor=south}, 
  	title=$\errorunif$, 
  	yticklabels={},
  ]
	
    \addplot
      table[x=Q_by_U,y expr=\thisrow{L_unif_bar}+1e-5,y error=L_unif_std] {\datauniform};
   
	\addplot
      table[x=Q_by_U,y expr=\thisrow{L_unif_bar}+1e-5,y error=L_unif_std] {\dataonline};
	
    \addplot
      table[x=Q_by_U,y expr=\thisrow{L_unif_bar}+1e-5,y error=L_unif_std] {\dataactive};

  \end{groupplot}
\end{tikzpicture} 

%% file: countermeasures.tex
\section{Extraction Countermeasures}
\label{sec:countermeasures}

We have shown in Sections~\ref{sec:withconf} and \ref{sec:real-world} that adversarial clients can effectively
extract ML models given access to rich prediction APIs. 
Given that this undermines
the financial models targeted by some ML cloud services, and potentially leaks
confidential training data, we believe researchers should
seek countermeasures. 

In \secref{sec:labelonly}, we analyzed the most obvious defense against our attacks: 
prediction API minimization. 
The constraint here is that the resulting API must still be useful in (honest) applications. 
For example, it is simple to change
APIs to not return confidences and not respond to incomplete queries, 
assuming applications can get by without it. This will prevent many of our
attacks, most notably the ones described in \secref{sec:withconf} as well as the feature discovery techniques used in our
Amazon case study (\secref{sec:real-world}). Yet, we showed that even if we strip an API to only provide class labels, successful attacks remain possible (\secref{sec:labelonly}), albeit at a much higher query cost.

We discuss further potential countermeasures below.

\paragraph{Rounding confidences.} \label{sec:rounding} Applications might need
confidences, but only at lower granularity. 
A possible defense is to round confidence scores to some
fixed precision~\cite{fredrikson15}. We note that ML APIs already work with
some finite precision when answering queries. For instance,
BigML reports confidences with $5$ decimal places, and Amazon provides values
with $16$ significant digits.

To understand the effects of limiting precision further, we 
re-evaluate equation-solving and decision tree path-finding attacks 
with confidence scores rounded to a
fixed decimal place. For equation-solving attacks, rounding the class
probabilities means that the solution to the obtained equation-system
might not be the target $f$, but some truncated version of it. For decision trees,
rounding confidence scores increases the chance of node $\id$
collisions, and thus decreases our attacks' success rate.

\begin{figure}[t]
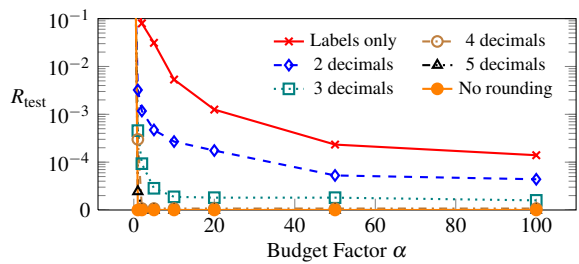

   \footnotesize
   \centering
   \include{data/regression/rounding/full_plot}
   \vspace{-10pt}
   \caption{\footnotesize{\textbf{Effect of rounding on model extraction.} Shows the average test error of equation-solving attacks on softmax models trained on the benchmark suite (\tabref{tab:benchmark}), as we vary the number of significant digits in reported class probabilities. Extraction with no rounding and with class labels only (adaptive retraining) are added for comparison.}}
   \label{fig:rounding:regression}%
\end{figure}

\figref{fig:rounding:regression} shows the results of experiments on softmax models, with class probabilities rounded to
$2$--$5$ decimals. We plot only $\errortest$, the results for
$\errorunif$ being similar. We observe that class probabilities
rounded to $4$ or $5$ decimal places (as done already in BigML) have no 
effect on the attack's success. When rounding further to $3$
and $2$ decimal places, the attack is weakened, but still vastly
outperforms adaptive retraining using class labels only.

For regression trees, rounding has no effect on our
 attacks. Indeed, for the models we considered, the output itself
is unique in each leaf (we could also round outputs, but
the impact on utility may be more critical).
For classification trees, we re-evaluated
our top-down attack, with confidence scores rounded to
fewer than $5$ decimal places. The attacks on the `IRS Tax Patterns' and `Email Importance' models are
the most resilient, and suffer no success degradation before scores
are rounded to $2$ decimal places. For the other models, rounding confidences to $3$ or $4$ decimal places severely undermines our attack.

\paragraph{Differential privacy.}
\label{sec:dp}
Differential privacy (DP)~\cite{dwork06} and its variants~\cite{li2013} have been explored as mechanisms for
protecting, in particular, the privacy of ML training data~\cite{pph}.
DP learning has been applied to regressions~\cite{functional-mech, chaudhuri2009privacy},
SVMs~\cite{rubinstein2009learning}, 
decision trees~\cite{jagannathan2009practical} and neural networks~\cite{shokri2015privacy}.
As some of our extraction attacks leak training data information (\secref{sec:withconf:klr}), one may ask whether
DP can prevent extraction, or at least reduce the severity of the privacy violations
that extraction enables. 

Consider naïve application of DP to protect 
individual training data elements. This should, in theory, decrease the
ability of an adversary $\advA$ to learn information about training set elements, when
given access to prediction queries. One would not expect, however,
that this prevents model extraction, as DP is not defined to do
so: consider a trivially useless learning algorithm for binary logistic regression,
that discards the training data
and sets $\wvec$ and $\beta$ to $0$. This algorithm
is differentially private, yet $\wvec$ and $\beta$ can easily be recovered 
using equation-solving.

A more appropriate strategy would be to apply DP directly to the
model parameters, which would amount to saying that a query should
not allow $\advA$ to distinguish between closely neighboring model
parameters. How exactly this would work and what privacy budgets would be
required is left as an open question by our work.

\paragraph{Ensemble methods.}
\label{sec:ensembles}
Ensemble methods such as random forests return as prediction an aggregation of
predictions by a number of
individual models. While we have not
experimented with ensemble methods as targets, we suspect that they
may be more resilient to extraction attacks, in the sense that
attackers will only be able to obtain relatively coarse approximations of the
target function. 
Nevertheless, ensemble methods may still be
vulnerable to other attacks such as model evasion~\cite{pdfrate14}.

%% file: data/regression/rounding/full_plot.tex
\pgfplotstableread[col sep=comma]{data/regression/rounding/extr_passive.dat}\data
\pgfplotstableread[col sep=comma]{data/regression/rounding/extr_passive_r5.dat}\datarfive
\pgfplotstableread[col sep=comma]{data/regression/rounding/extr_passive_r4.dat}\datarfour
\pgfplotstableread[col sep=comma]{data/regression/rounding/extr_passive_r3.dat}\datarthree
\pgfplotstableread[col sep=comma]{data/regression/rounding/extr_passive_r2.dat}\datartwo
\pgfplotstableread[col sep=comma]{data/regression/multiclass/softmax/base_adapt-local.dat}\dataactive

\begin{tikzpicture}
  \begin{groupplot}[group style={group size=2 by 1, horizontal sep = 30pt},cycle list name=list]

  \nextgroupplot[
  	  ymode=log, 
  	  xlabel=Budget Factor $\alpha$, 
  	  xlabel style={at={(axis description cs:0.5,-0.1)},anchor=south}, 
  	  ylabel=$\errortest$, 
  	  ymin=1e-5,
  	  ymax=0.1,
  	  xtick pos=left,
  	  ytick={1e-5, 1e-4, 1e-3, 1e-2, 1e-1},
  	  width=\columnwidth,
  	  height=0.52 \columnwidth,
  	  log origin=infty,
  	  yticklabels={$0$, $10^{-4}$, $10^{-3}$, $10^{-2}$, $10^{-1}$},
  	  ylabel style={at={(axis description cs:0.05,.5)},rotate=-90,anchor=south}, 
  	  legend columns=3,
  	  transpose legend,
  	  legend style = {font=\scriptsize, draw = none, fill=none, 
  	  				row sep=-1pt,
  	  				 /tikz/every even column/.append style={column sep=0.25cm},
  	  				 legend image code/.code={
  	      			 	\draw[mark repeat=2,mark phase=2]
						plot coordinates {
							(0cm,0cm)
							(0.2cm,0cm)        
							(0.4cm,0cm)         
						};}
					},
    ]
	
    \addplot
      table[x=Q_by_U,y expr=\thisrow{L_test_bar}+1e-5,y error=L_test_std, col sep=comma] {\dataactive};
    \addlegendentry{Labels only};	
	
	\addplot
      table[x=Q_by_U,y expr=\thisrow{L_test_bar}+1e-5,y error=L_test_std, col sep=comma] {\datartwo};
    \addlegendentry{2 decimals};	
    
    \addplot
      table[x=Q_by_U,y expr=\thisrow{L_test_bar}+1e-5,y error=L_test_std, col sep=comma] {\datarthree};
    \addlegendentry{3 decimals};	
    
    \addplot
      table[x=Q_by_U,y expr=\thisrow{L_test_bar}+1e-5,y error=L_test_std, col sep=comma] {\datarfour};
    \addlegendentry{4 decimals};
    	
    \addplot
      table[x=Q_by_U,y expr=\thisrow{L_test_bar}+1e-5,y error=L_test_std, col sep=comma] {\datarfive};
    \addlegendentry{5 decimals};
    
    \addplot
      table[x=Q_by_U,y expr=\thisrow{L_test_bar}+1e-5,y error=L_test_std, col sep=comma] {\data};
    	\addlegendentry{No rounding};

  \end{groupplot}    
\end{tikzpicture} 

%% file: relwork.tex
\section{Related Work}
\label{sec:relwork}

Our work is related to the extensive literature on learning theory, such as PAC learning~\cite{valiant84} and its
variants~\cite{angluin1988queries,benedek1991learnability}. Indeed, extraction can be viewed as a type
of learning, in which an unknown instance of a known hypothesis class (model
type) is providing labels (without error).  This is often called learning with
membership queries~\cite{angluin1988queries}. Our setting differs from these in two 
ways. The first is conceptual: in PAC learning one builds algorithms to learn a
concept --- the terminology belies the motivation of formalizing 
learning from data. In model extraction,
an attacker is literally given a function oracle that it seeks to illicitly
determine.  
The second difference is
more pragmatic: prediction APIs reveal richer information than
assumed in prior learning theory work, and we exploit that. 

Algorithms for learning with membership queries have been proposed for  Boolean functions~\cite{bshouty1995exact,jackson1994efficient,kushilevitz1993learning, bellare1992technique} and various binary classifiers~\cite{lowd2005adversarial,stevens2013hardness,nelson2012query}. The latter line of work, initiated by Lowd and Meek~\cite{lowd2005adversarial}, studies strategies for model evasion, in the context of spam or fraud detectors~\cite{biggio2013evasion,pdfrate14,huang2011adversarial,lowd2005adversarial,lowd2005good}. 
Intuitively, model extraction seems harder than evasion, and this is corroborated by results from theory~\cite{nelson2012query,stevens2013hardness,lowd2005adversarial} and practice~\cite{pdfrate14,lowd2005adversarial}.

Evasion attacks fall into the larger field of \emph{adversarial machine learning}, that studies machine learning in general adversarial settings~\cite{barreno2006can,huang2011adversarial}. In that context, a number of authors have considered strategies and defenses for \emph{poisoning} attacks, that consist in injecting maliciously crafted samples into a model's train or test data, so as to decrease the learned model's accuracy~\cite{dalvi2004adversarial,biggio2012poisoning,
kloft2010online,newsome2006paragraph,rubinstein2009antidote}.

In a concurrent work, Papernot et al.~\cite{papernot2016practical}
make use of improper model extraction techniques in the context of model evasion
attacks against deep neural networks.
They first extract a ``substitute'' network using data-augmentation techniques 
similar in spirit to our adaptive retraining strategy (Section~\ref{sec:labelonly}). 
This
substitute is then in turn used to craft adversarial samples likely to fool the
target black-box model. For their attacks to remain tractable despite the
complexity of the targeted models, they aim for lower extraction accuracy
($\errortest \approx 80\%$), yet show that this suffices for adversarial samples
to ``transfer'' from one model to the other with high probability.

In a non-adversarial setting, improper model extraction techniques have been 
applied for interpreting~\cite{craven1996extracting,towell1993extracting,
andrews1995survey} and compressing~\cite{model_compression, distilling} 
complex neural networks. 

%% file: conclusion.tex
\section{Conclusion}
\label{sec:conclusion}

We demonstrated how the flexible prediction APIs exposed by current ML-as-a-service providers enable new model extraction attacks that could subvert model monetization, violate training-data privacy, and facilitate model evasion.
Through local experiments and online attacks on two major providers, BigML and Amazon, we illustrated the efficiency and broad applicability of attacks that exploit common API features, such as the availability of confidence scores or the ability to query arbitrary partial inputs. We presented a generic \emph{equation-solving} attack for models with a logistic output layer and a novel \emph{path-finding} algorithm for decision trees.

We further explored potential countermeasures to these attacks, the most obvious being a restriction on the information provided by ML APIs. Building upon prior work from learning-theory, we showed how an attacker that only obtains class labels for adaptively chosen inputs, may launch less effective, yet potentially harmful, \emph{retraining attacks}.
Evaluating these attacks, as well as more refined countermeasures, on production-grade ML services is an interesting avenue for future work.

\smallskip
\noindent\textbf{Acknowledgments.}\;
We thank Mart\'{i}n Abadi and the anonymous reviewers for their
comments.  This work was supported by NSF grants 
1330599, 1330308, and 1546033, as well as a generous gift
from Microsoft.

%% file: model-details.tex
\section{Some Details on Models}
\label{sec:model-details} 

\textbf{SVMs.}\;
Support vector machines (SVMs) perform binary classification
($\numclasses = 2$)
by defining a maximally separating hyperplane in $\fdim$-dimensional feature
space. A linear SVM is a function 
$f(\xvec) = \sign(\wvec \cdot \xvec + \beta)$ where `$\sign$' outputs 0 for all
negative inputs and 1 otherwise. Linear SVMs are
not suitable for non-linearly separable data. Here one uses instead kernel
techniques~\cite{boser1992training}. 

A kernel is a function $\kernel\colon\fspace\times\fspace\rightarrow\R$. Typical
kernels include the quadratic kernel $\kernelquad(\xvec,\xvec') = (\xvec^T \cdot
\xvec' + 1)^2$ and the Gaussian radial basis function (RBF) kernel
$\kernelrbf(\xvec,\xvec') = e^{-\gamma ||\xvec - \xvec'||^2}$, parameterized by a value $\gamma \in \R$. 
A kernel's projection function is a map $\featext$
defined by $\kernel(\xvec,\xvec') = \featext(\xvec) \cdot \featext(\xvec')$.  We do not
use $\featext$ explicitly, indeed for RBF kernels this produces an
infinite-dimension vector. Instead, 
classification is defined using a ``kernel trick'': 
$f(\xvec) = \sign(\left[\sum_{i=1}^t
\alpha_i\kernel(\xvec,\xvec_i)\right] + \beta)$ where $\beta$ is again a learned threshold, $\alpha_1,\ldots,\alpha_t$ are
learned weights, and $\xvec_1,\ldots,\xvec_t$ are feature
vectors of inputs from a training set. The $\xvec_i$ for which $\alpha_i \ne 0$
are called support vectors. Note that for non-zero $\alpha_i$, it is
the case that $\alpha_i < 0$ if
the training-set label of $\xvec_i$ was zero and $\alpha_i > 0$ otherwise.

\smallskip
\noindent\textbf{Logistic regression.}\;
SVMs do not directly generalize to
multiclass settings $\numclasses > 2$, nor do they output class probabilities. Logistic regression (LR) is a popular
classifier that does. A binary LR model
is defined as $f_1(\xvec) = \sigma(\wvec\cdot \xvec + \beta) = 1/(1+e^{-(\wvec \cdot
\xvec + \beta)})$ and $f_0(\xvec) = 1-f_1(\xvec)$. A class label is chosen as 1 iff $f_1(\xvec) > 0.5$.

When $\numclasses > 2$, one fixes $\numclasses$ weight vectors
$\wvec_0,\ldots,\wvec_{\numclasses-1}$ each in $\R^\fdim$, thresholds
$\beta_0,\ldots,\beta_{\numclasses-1}$ in $\R$  and defines 
  $f_i(\xvec) = e^{\wvec_i\cdot \xvec + \beta_i} / (
  \sum_{j=0}^{\numclasses-1} e^{\wvec_j\cdot \xvec + \beta_j})$
for $i \in \Ints_\numclasses$.
The class label is taken to be $\argmax_i f_i(\xvec)$. Multiclass regression is referred to as multinomial or softmax regression.
An alternative approach to softmax regression is to build a binary model $\sigma(\wvec_i\cdot \xvec + \beta_i)$ per class in a \emph{one-vs-rest} fashion and then set $f_i(\xvec) = {\sigma(\wvec_i\cdot \xvec + \beta_i)}/{\sum_j \sigma(\wvec_j\cdot \xvec + \beta_j)}$.

These are log-linear models, and may not be suitable for data that is not
linearly separable in $\fspace$. Again, one may use kernel techniques to deal with
more complex data relationships (c.f.,~\cite{zhu01}). Then, one replaces
$\wvec_i\cdot\xvec + \beta_i$ with $\sum_{r=1}^t \alpha_{i,r}
\kernel(\xvec,\xvec_r) + \beta_i$. As written, this uses the entire set of
training data points $\xvec_1,\ldots,\xvec_t$ as so-called representors (here
analogous to support vectors).  Unlike with SVMs, where most training data set
points will never end up as support vectors, here all training set points are
potentially representors. In practice one uses a size $s < t$ random subset
of training data~\cite{zhu01}.

\smallskip
\noindent\textbf{Deep neural networks.}\;
A popular way of extending softmax regression to handle data that is non linearly separable in $\fspace$ is to first apply one or more non-linear transformations to the input data. The goal of these \emph{hidden layers} is to map the input data into a (typically) lower-dimensional space in which the classes are separable by the softmax layer. We focus here on fully connected networks, also known as multilayer perceptrons, with a single hidden layer. The hidden layer consists of a number $h$ of \emph{hidden nodes}, with associated weight vectors $\wvec^{(1)}_0, \ldots, \wvec^{(1)}_{h-1}$ in $\Reals^\fdim$ and thresholds $\beta^{(1)}_0, \ldots, \beta^{(1)}_{h-1}$ in $\R$. The $i$-th hidden unit applies a non linear transformation $h_i(\xvec) = g(\wvec^{(1)}_i\cdot \xvec + \beta^{(1)}_i)$, where $g$ is an activation function such as $\tanh$ or $\sigma$. The vector $h(\xvec) \in \Reals^h$ is then input into a softmax output layer with weight vectors $\wvec^{(2)}_0, \ldots, \wvec^{(2)}_{\numclasses-1}$ in $\Reals^h$ and thresholds $\beta^{(2)}_0, \ldots, \beta^{(2)}_{\numclasses-1}$ in $\R$.

\smallskip
\noindent\textbf{Decision trees.}\;
A decision tree $\tree$ is a labeled tree. Each internal node $v$ is labeled by a feature index
$i \in \set{1, \dots, \fdim}$ and a \emph{splitting function} $\splitfun: \fspace_i
\rightarrow \Ints_{k_v}$, where $k_v \geq 2$ denotes the number of outgoing
edges of $v$.

On an input $\xvec = (x_1, x_2, \dots, x_\fdim)$, a tree $T$ defines a computation
as follows, starting at the root. When we reach a
node $v$, labeled by $\set{i, \splitfun}$, we proceed to the child of $v$
indexed by $\splitfun(x_i)$. We consider three types of splitting functions
$\splitfun$
that are typically used in practice (\cite{bigml}): 
\begin{newenum} 
\item The feature $x_i$
is categorical with $\fspace_i = \Ints_k$. Let $\set{S, T}$ be some partition of
$\Ints_k$. Then $k_v = 2$ and $\splitfun(x_i) = 0$ if $x_i \in S$ and $\splitfun(x_i)=1$
if $x_i \in T$. This is a binary split on a categorical feature.

\item The feature $x_i$ is categorical with $\fspace_i = \Ints_k$. We have $k_v
= k$ and $\splitfun(x_i) = x_i$. This corresponds to a $k$-ary split on a categorical
feature of arity $k$.

\item The feature $x_i$ is continuous with $\fspace_i = [a,b]$. Let $a<t<b$ be
a \emph{threshold}. Then $k_v = 2$ and $\splitfun(x_i) = 0$ if $x_i \leq t$ and
$\splitfun(x_i)=1$ if $x_i > t$. This is a binary split on a continuous
feature with threshold $t$.
\end{newenum}
When we reach a leaf, we terminate and output that leaf's value. This value can be a class label, or a class label and confidence score. This defines a function $f:\fspace\rightarrow \range$.

%% file: datasets.tex
\section{Details on Data Sets}
\label{sec:datasets}

Here we give some more information about the data sets we used
in this work. Refer back to \tabref{tab:benchmark} and
\tabref{tab:benchmark-trees}.  

\smallskip
\noindent\textbf{Synthetic data sets.}\;
We used 4 synthetic data sets from \texttt{scikit}~\cite{scikit}. 
The first two data sets are classic examples of non-linearly separable
data, consisting of two concentric \emph{Circles}, or two interleaving
\emph{Moons}. The next two synthetic data sets, \emph{Blobs} and
\emph{5-Class}, consist of Gaussian clusters of points assigned to
either $3$ or $5$ classes. 

\smallskip
\noindent\textbf{Public data sets.}\;
We gathered a varied set of data sets representative of the type of data we would expect ML service users to use to train logistic and SVM based models. These include famous data sets used for supervised learning, obtained from the UCI ML repository (\emph{Adult}, \emph{Iris}, \emph{Breast Cancer}, \emph{Mushrooms}, \emph{Diabetes}). We also consider the \emph{Steak} and \emph{GSS} data sets used in prior work on model inversion~\cite{fredrikson15}. Finally, we add a data set of digits available in \texttt{scikit}, to visually illustrate training data leakage in kernelized logistic models (c.f. \secref{sec:withconf:klr}).

\smallskip
\noindent\textbf{Public data sets and models from BigML.}\;
For experiments on decision trees, we chose a varied set of models publicly available on BigML's platform. These models were trained by real MLaaS users and they cover a wide range of application scenarios, thus providing a realistic benchmark for the evaluation of our extraction attacks.

The \emph{IRS} model predicts a US state, based on administrative tax records.
The \emph{Steak} and \emph{GSS} models respectively predict a person's preferred steak
preparation and happiness level, from survey and demographic data. These two models were also considered in~\cite{fredrikson15}. The
\emph{Email Importance} model predicts whether Gmail classifies an email as `important'
or not, given message metadata. The \emph{Email Spam} model classifies emails as spam,
given the presence of certain words in its content. 
The German Credit data set was taken from the UCI library~\cite{uci} and classifies a user's loan risk.
Finally, two regression models respectively predict \emph{Medical
Charges} in the US based on state demographics, and the \emph{Bitcoin Market
Price} from daily opening and closing values.

%% file: trees-analysis.tex
\section{Analysis of the Path-Finding Algorithm}
\label{sec:trees-analysis}

In this section, we analyze the correctness and complexity of the decision tree extraction algorithm in \algref{alg:extract}. We assume that all leaves are assigned a unique $\id$ by the oracle $\oracle$, and that no continuous feature is split into intervals of width smaller than $\epsilon$. We may use $\id$ to refer directly to the leaf with identity $\id$.

\smallskip
\noindent\textbf{Correctness.}\;
Termination of the algorithm follows immediately from the fact that new queries are only added to $Q$ when a new leaf is visited. As the number of leaves in the tree is bounded, the algorithm must terminate.

We prove by contradiction that all leaves are eventually visited. Let the \emph{depth} of a node $v$, denote the length of the path from $v$ to the root (the root has depth $0$). For two leaves $\id, \id'$, let $A$ be their deepest common ancestor ($A$ is the deepest node appearing on both the paths of $\id$ and $\id'$). We denote the depth of $A$ as $\Delta(\id, \id')$.

Suppose \algref{alg:extract} terminates without visiting all leaves, and let $(\id, \id')$ be a pair of leaves with maximal  $\Delta(\id, \id')$, such that $\id$ was visited but $\id'$ was not. Let $x_i$ be the feature that their deepest common ancestor $A$ splits on. When $\id$ is visited, the algorithm calls \texttt{LINE\_SEARCH} or \texttt{CATEGORY\_SPLIT} on feature $x_i$. As all leaf $\id$s are unique and there are no intervals smaller than $\epsilon$, we will discover a leaf in each sub-tree rooted at $A$, including the one that contains $\id'$. Thus, we visit a leaf $\id''$ for which $\Delta(\id'', \id') > \Delta(\id, \id')$, a contradiction.

\smallskip
\noindent\textbf{Complexity.}\;
Let $m$ denote the number of leaves in the tree. Each leaf is visited exactly once, and for each leaf we check all $\fdim$ features. Suppose continuous features have range $[0,b]$, and categorical features have arity $k$.
For continuous features, finding one threshold takes at most $\log_2(\frac{b}\epsilon)$ queries. 
As the total number of splits on one feature is at most $m$ (\ie all nodes split on the same feature), finding all thresholds uses at most $m \cdot \log_2(\frac b \epsilon)$ queries. 
Testing a categorical feature uses $k$ queries.
The total query complexity is $O(m \cdot (\fdim_{cat} \cdot k + \fdim_{cont}\cdot m \cdot\log(\frac{b}{\epsilon}))$, where $\fdim_{cat}$ and $\fdim_{cont}$ represent respectively the number of categorical and continuous features.

For the special case of boolean trees, the complexity is $O(m\cdot d)$. In comparison, the algorithm of~\cite{kushilevitz1993learning}, that uses membership queries only, has a complexity polynomial in $\fdim$ and $2^\delta$, where $\delta$ is the tree depth. For degenerate trees, $2^\delta$ can be exponential in $m$, implying that the assumption of unique leaf identities (obtained from confidence scores for instance) provides an exponential speed-up over the best-known approach with class labels only. The algorithm from~\cite{kushilevitz1993learning} can be extended to regression trees, with a complexity polynomial in the size of the output range $\outspace$. Again, under the assumption of unique leaf identities (which could be obtained solely from the output values) we obtain a much more efficient algorithm, with a complexity independent of the output range.

\smallskip
\noindent\textbf{The Top-Down Approach.}\;
The correctness and complexity of the top-down algorithm from \secref{sec:dectrees} (which uses incomplete queries), follow from a similar analysis. The main difference is that we assume that all nodes have a unique $\id$, rather than only the leaves.

%% file: improper.tex
\section{A Note on Improper Extraction}
\label{sec:improper}

To extract a model $f$, without knowledge of the model class, a simple strategy is to extract a multilayer perceptron $\fout$ with a large enough hidden layer.
Indeed, feed-forward networks with a single hidden layer can, in principle, closely approximate any continuous function over a bounded subset of $\R^\fdim$~\cite{hornik1989multilayer,cybenko1989approximation}.

However, this strategy intuitively does not appear to be optimal. Even if we know that we can find a multilayer perceptron $\fout$ that closely matches $f$, $\fout$ might have a far more complex representation (more parameters) than $f$. Thus, tailoring the extraction to the `simpler' model class of the target $f$ appears more efficient. In learning theory, the problem of finding a succinct representation of some target model $f$ is known as \emph{Occam Learning}~\cite{blumer1990occam}.

Our experiments indicate that such generic improper extraction indeed appears sub-optimal, in the context of equation-solving attacks. We train a softmax regression over the Adult data set with target ``Race''. The model $f$ is defined by $530$ real-valued parameters. As shown in \secref{sec:withconf:multiclass}, using only $530$ queries, we extract a model $\fout$ from the \emph{same} model class, that closely matches $f$ ($\fout$ and $f$ predict the same labels on 100\% of tested inputs, and produce class probabilities that differ by less than $10^{-7}$ in TV distance). We also extracted the same model, assuming a multilayer perceptron target class. Even with $1{,}000$ hidden nodes (this model has $111{,}005$ parameters), and 10$\times$ more queries ($5{,}300$), the extracted model $\fout$ is a weaker approximation of $f$ ($99.5\%$ accuracy for class labels and TV distance of $10^{-2}$ for class probabilities).